\documentclass[reprint,aps,pra,showpacs]{revtex4-1}

\usepackage{amsmath,graphicx,dcolumn,hyperref,graphicx,comment}
\usepackage{hyperref}
\hypersetup{colorlinks=true,linkcolor=blue,citecolor=blue,urlcolor=blue}
\usepackage{newtxtext,newtxmath}
\usepackage[T1]{fontenc}
\usepackage[utf8]{inputenc}

\let\originalleft\left
\let\originalright\right
\renewcommand{\left}{\mathopen{}\mathclose\bgroup\originalleft}
\renewcommand{\right}{\aftergroup\egroup\originalright}

\renewcommand{\vec}[1]{\mathbf{#1}}

\begin{document}
\frenchspacing

\title{Calculations of positronium-atom scattering using a spherical cavity}
\author{A. R. Swann}\email{a.swann@qub.ac.uk}
\author{G. F. Gribakin}\email{g.gribakin@qub.ac.uk}
\affiliation{School of Mathematics and Physics, Queen's University Belfast, Belfast BT7 1NN, United Kingdom}

\date{\today}

\begin{abstract}
Positronium (Ps) scattering by noble-gas atoms (He, Ne, Ar, Kr, and Xe) is studied in the frozen-target approximation and with inclusion of the van der Waals interaction. Single-particle electron and positron states in the field of the target atom are calculated, with the system enclosed by a hard spherical wall. The two-particle Ps wave function is expanded in these states, and the Hamiltonian matrix is diagonalized, giving the Ps energy levels in the cavity. Scattering phase shifts, scattering lengths, and cross sections are extracted from these energies and compared with existing calculations and experimental data. Analysis of the effect of the van der Waals interaction shows that it cannot explain the recent experimental data of Brawley \textit{et al.} for Ar and Xe [Phys. Rev. Lett. \href{http://dx.doi.org/10.1103/PhysRevLett.115.223201}{\textbf{115}}, \href{http://dx.doi.org/10.1103/PhysRevLett.115.223201}{223201} (\href{http://dx.doi.org/10.1103/PhysRevLett.115.223201}{2015})].
\end{abstract}


\maketitle

\section{\label{sec:intro}Introduction}

In this paper a new theoretical method for the calculation of positronium-atom scattering is developed. It involves confining the electron-positron pair to a large sphere with the atom placed at the centre. The scattering problem is thus reduced to the calculation of bound states. This enables us to obtain converged results that fully allow for the distortion of positronium. This is also an important step towards the development of a many-body-theory approach to the problem.

The interaction of positronium (Ps) with matter and antimatter is an interesting topic \cite{Laricchia12} with applications in many areas of physics. For example, the proposed AEgIS and GBAR experiments at CERN \cite{Kellerbauer08,Debu2012} aim to test whether gravity affects antimatter in the same way as matter, making antihydrogen in Ps collisions with antiprotons, with Ps produced in a mesoporous material \cite{Consolati13}. Ps is widely utilized in condensed-matter physics to determine pore sizes in nanoporous materials and to probe intermolecular voids in polymers \cite{Gidley06}. Moreover, Ps formation in porous materials is used to study its interactions with gases, e.g., Xe \cite{Shibuya13,Shibuya13a}, or the interaction between the Ps atoms themselves \cite{Cassidy07,Cassidy08,Cassidy11,Cassidy12}, with prospects of room-temperature Bose-Einstein condensation and an annihilation gamma laser \cite{Platzman94,Mills07,Cassidy07a}. There are also proposals for using a beam of long-lived Rydberg Ps for measuring the free fall of a matter-antimatter system \cite{Cassidy14} and for detecting positron-atom bound states \cite{Swann16}.

In this work we consider low-energy Ps scattering from noble-gas atoms. Experiments found that for energies above the Ps breakup threshold (6.8~eV), the total Ps scattering cross sections for atoms and small molecules are similar to those of electrons of equal velocity \cite{Brawley10,Brawley12}, as though the positron played no part in the scattering. An explanation of this phenomenon was obtained using the impulse approximation \cite{Fabrikant14a}. Recent measurements for Ar and Xe indicate 
that at low energies the Ps-atom scattering cross section becomes very small, suggesting that a Ramsauer minimum may be present \cite{Brawley15}. So far this behavior has not been reproduced by the calculations \cite{Fabrikant14,Gribakin16}, and one of the aims of our paper is to subject this phenomenon to a closer scrutiny. 

Besides the experiments that use a Ps beam \cite{Zafar96}, low-energy Ps-atom scattering is studied indirectly through positron annihilation in gases, where it determines the rate of Ps thermalization. The latter is important for precision measurements that test QED in this purely leptonic system \cite{Vallery03}. However, there is little accord among the published data on the Ps momentum-transfer cross sections (see below), and we hope that the present calculations can provide some certainty here.

Ps-atom scattering has been a difficult subject to research theoretically, since both the target and projectile have an internal structure. The earliest calculation was carried out in 1954 by Massey and Mohr \cite{Massey54}; they studied Ps-H collisions in the first Born approximation and in the Born-Oppenheimer approximation, and emphasized the importance of electron exchange. 
With particular regard to the noble gases, there is a wealth of theoretical calculations for low-energy Ps scattering from helium. Fraser \cite{Fraser62,Fraser68} and Fraser and Kraidy \cite{Fraser66} initiated the study of Ps-He scattering using a static-exchange approach, with both the Ps and He atoms ``frozen'' in their ground states. Barker and Bransden \cite{Barker68,Barker69} considered the problem both in the static-exchange approximation and including the adiabatic part of the van der Waals potential. Drachman and Houston \cite{Drachman70} estimated the scattering length and pickoff annihilation rate using a model potential.
Interest in the subject waned after 1970, until McAlinden \textit{et al.} revisited it in 1996 \cite{McAlinden96}, using the target-elastic pseudostate close-coupling method and neglecting electron exchange between the Ps and the atom. Shortly thereafter, Sarkar and Ghosh \cite{Sarkar97} presented scattering phase shifts and cross sections found using a static-exchange model. The year 1999 saw much development on the topic: Biswas and Adhikari \cite{Biswas99} and Sarkar \textit{et al.} \cite{Sarkar99} performed three-state [Ps($1s$,$2s$,$2p$)-He($1s^2$)] close-coupling calculations, while Blackwood \textit{et al.} \cite{Blackwood99} improved upon the work of McAlinden \textit{et al.} \cite{McAlinden96} by accounting for exchange. Further calculations using the close-coupling method appeared within the next two years \cite{Adhikari00,Ghosh01,Basu01}. From 2001, calculations employing variational methods began to appear in the literature: Adhikari \cite{Adhikari01} considered $S$-wave Ps-He scattering and obtained an estimate for the scattering length, while Ivanov \textit{et al.} \cite{Ivanov01} and Mitroy and Ivanov \cite{Mitroy01} used the fixed-core stochastic-variational method with one- and two-body polarization potentials to find low-energy $S$-wave phase shifts and scattering lengths. Chiesa \textit{et al.} \cite{Chiesa02} applied the diffusion Monte Carlo method to compute phase shifts and threshold cross sections. In 2003, DiRenzi and Drachman \cite{DiRenzi03} noted the vast discrepancy between the earlier work  \cite{Drachman70} and Ref.~\cite{Blackwood99} and undertook a reexamination of the former work; they concluded that an assumption made therein (that the direct potential was negligible in comparison to the exchange potential) was not quantitatively correct. A 2004 work by Walters \textit{et al.} \cite{Walters04} presented new results in the pseudostate-close-coupling framework, now allowing for excitations of the target He atom in addition to the Ps atom. Most recently, DiRenzi and Drachman \cite{DiRenzi13} calculated the Ps-He scattering length using a variational wave function.

Calculations of low-energy Ps scattering on the other noble-gas atoms (Ne, Ar, Kr, and Xe) are much more sparse. The earliest calculations we are aware of were by McAlinden \textit{et al.} \cite{McAlinden96}, who considered Ps scattering on Ar (in addition to H and He) using the pseudostate-close-coupling method, neglecting the effects of electron exchange \cite{McAlinden96}.
Biswas and Adhikari \cite{Biswas00} similarly used the coupled-channel framework to investigate Ps scattering from H, He, Ne, and Ar. In addition to Ps-He scattering, Mitroy and Ivanov \cite{Mitroy01} applied the fixed-core stochastic-variational method to study Ps scattering from Ne, Ar, Li$^+$, Na$^+$, and K$^+$; Mitroy and Bromley \cite{Mitroy03} extended this work to scattering from Kr and Xe. Blackwood \textit{et al.} \cite{Blackwood02,Blackwood03} developed a coupled-state treatment for Ps scattering from Ne, Ar, Kr, and Xe in the frozen-target approximation; for Kr and Xe the incident Ps was also frozen in its ground state. Finally, Fabrikant and Gribakin \cite{Fabrikant14} and Gribakin \textit{et al.} \cite{Gribakin16} employed a pseudopotential method to study low-energy Ps collisions with Ar, Kr, and Xe.

On the experimental front, there were several studies of Ps scattering on the noble gases \cite{Canter75,Rytsola84,Coleman94,Nagashima95,Garner96,Skalsey98,Nagashima98,Garner98,Nagashima01,Skalsey03,Engbrecht08,Brawley10,Sano15}. Most of these determined momentum-transfer cross sections at energies below the Ps ionization potential (6.8~eV), and some inferred zero-energy elastic cross sections. Direct measurements of the elastic cross section for energies below $6.8$~eV have only been carried out recently, and only for Ar and Xe \cite{Brawley15}.

In this work we present new theoretical calculations of Ps scattering on He, Ne, Ar, Kr, and Xe at energies below the Ps ionization potential. A $B$-spline basis \cite{deBoor01,Sapirstein96,Bachau01} is used to construct states of Ps in the field of the target atom, and the entire system is enclosed in an impenetrable spherical cavity. Apart from the presence of the atom, most details of the calculation are similar to our earlier study of Ps states in an empty spherical cavity \cite{Brown17}. Here we initially use the frozen-target approximation, where distortion of the Ps is accounted for, but the target atom is ``frozen'' in its ground state. Subsequently, the long-range van der Waals interaction between the target and projectile is included using a model potential. Scattering phase shifts, scattering lengths, and elastic and momentum-transfer cross sections are recorded, and comparisons are made with existing calculations and experimental data.

Unless otherwise stated, atomic units (a.u.) are used throughout; the symbol $a_0$ denotes the Bohr radius (the atomic unit of length).

\section{\label{sec:theory}Theory}

\subsection{Frozen-target approximation}

The electron density and electrostatic potential of a noble-gas atom in the ground state are described well by the Hartree-Fock approximation. The total Hamiltonian for Ps in the static Hartree-Fock field of the atom placed at the origin is
\begin{equation}\label{eq:H}
H = -\frac{1}{2} \nabla_e^2 - \frac{1}{2} \nabla_p^2 + U_e(\mathbf{r}_e) + U_p(\mathbf{r}_p) + V(\mathbf{r}_e,\mathbf{r}_p),
\end{equation}
where $\mathbf{r}_e$ ($\mathbf{r}_p$) is the position of the electron (positron) in Ps relative to the nucleus of the target atom, $U_e$ ($U_p$) is the static potential of the target atom for the electron (positron) in Ps, and $V(\mathbf{r}_e,\mathbf{r}_p)=-\lvert \mathbf{r}_e - \mathbf{r}_p \rvert^{-1}$ is the Coulomb interaction between the electron and positron in Ps. The atomic potential $U_e$ is the sum of the direct and (nonlocal) exchange potentials, while $U_p$ consists only of the direct potential.

The entire system is enclosed by an impenetrable spherical wall of radius $R_c$ centered on the target atom; consequently, all Ps states are discrete \cite{Brown17}. This hard-wall cavity is a key feature of our method. Values of $R_c$ are chosen in such a way that the cavity does not affect the atomic ground state and allows for an accurate description of Ps ``bouncing'' off the atom (Sec.~\ref{sec:numimp}). This enables us to determine Ps-atom scattering phase shifts from the discrete energy eigenvalues (Sec.~\ref{sec:pha_scs}).

To solve the Schr\"odinger equation for the Hamiltonian (\ref{eq:H}), we first consider single-particle electron and positron states in the field of the atom; we denote these by $\varphi_\mu(\mathbf{r}_e)$ and $\varphi_\nu(\mathbf{r}_p)$ respectively. They satisfy the following equations:
\begin{subequations}
\begin{align}\label{eq:el}
\left[ -\frac{1}{2} \nabla_e^2 + U_e \right] \varphi_\mu(\mathbf{r}_e) &= \varepsilon_\mu \varphi_\mu(\mathbf{r}_e) , \\ \label{eq:pos}
\left[ -\frac{1}{2} \nabla_p^2 + U_p \right] \varphi_\nu(\mathbf{r}_p) &= \varepsilon_\nu \varphi_\nu(\mathbf{r}_p), 
\end{align}
\end{subequations}
where $\varepsilon_\mu$ and $\varepsilon_\nu$ are the electron and positron energy eigenvalues, and the indices $\mu$ and $\nu$ stand for the radial and angular momentum quantum numbers. Since $U_e$ and $U_p$ are spherically symmetric, Eqs. (\ref{eq:el}) and (\ref{eq:pos}) are reduced to radial equations by separating the angular parts of the wave functions, e.g., $\varphi_\mu(\vec{r}_e)=r_e^{-1}P_{\varepsilon l}(r_e)Y_{l m}(\Omega _e)$, and the radial wave functions are subject to the boundary condition $P_{\varepsilon l}(R_c)=0$.

Being eigenstates of the single-particle Hamiltonians, the sets of functions $\{\varphi_\mu\}$ and $\{\varphi_\nu\}$ are orthonormal. From these, a two-particle Ps wave function with fixed total angular momentum $J$ and parity $\Pi$ may be constructed as
\begin{equation}\label{eq:CIexpansion}
\Psi_{J\Pi}(\mathbf{r}_e,\mathbf{r}_p) = \sum_{\mu,\nu} C_{\mu\nu} \varphi_\mu(\mathbf{r}_e) \varphi_\nu(\mathbf{r}_p),
\end{equation}
where the $C_{\mu\nu}$ are expansion coefficients. In theory, the sum in Eq. (\ref{eq:CIexpansion}) should run over all orbital angular momenta and radial quantum numbers, up to infinity, but in practice we use finite maximum values $l_\text{max}$ and $n_\text{max}$, respectively (see Sec.~\ref{sec:numimp}).

Substitution of Eq.~(\ref{eq:CIexpansion}) into the Schr\"odinger equation
\begin{equation}
H \Psi_{J\Pi}=E\Psi_{J\Pi}
\end{equation}
leads to a matrix-eigenvalue equation
\begin{equation}
\mathbf{H}\mathbf{C}=E\mathbf{C}.
\end{equation}
Here, the Hamiltonian matrix $\mathbf{H}$ has elements
\begin{equation}\label{eq:ham_mat_elem}
\langle \nu' \mu' \vert H \vert \mu \nu \rangle = \left( \varepsilon_\mu + \varepsilon_\nu \right) \delta_{\mu\mu'} \delta_{\nu\nu'} + \langle \nu' \mu' \vert V \vert \mu \nu \rangle ,
\end{equation}
with the Coulomb matrix elements defined as
\begin{equation*}
\langle \nu' \mu' \vert V \vert \mu \nu \rangle =
-\iint  \frac{\varphi_{\nu'}^*(\mathbf{r}_p) \varphi_{\mu'}^*(\mathbf{r}_e) \varphi_{\mu}(\mathbf{r}_e) \varphi_{\nu}(\mathbf{r}_p)}{\lvert \mathbf{r}_e - \mathbf{r}_p \rvert}\,d^3\mathbf{r}_e\, d^3\mathbf{r}_p,
\end{equation*}
and $\mathbf{C}$ is the vector of the expansion coefficients $C_{\mu\nu}$. Diagonalization of the Hamiltonian matrix yields the energy eigenvalues $E$ and the expansion coefficients. With the electron and positron states separated into angular and radial parts, integration over the angular variables and coupling of the angular momenta in Eq.~(\ref{eq:CIexpansion}) are performed analytically (see Appendix~A of Ref. \cite{Brown17} for details).

We restrict our interest to collisions of Ps in the ground state ($1s$), which has an internal energy of $-1/4$~a.u. The total Ps energy for such states can be written as
\begin{equation}\label{eq:E}
E=-\frac{1}{4}+\frac{K^2}{2m},
\end{equation}
where $K$ is the Ps center-of-mass momentum and $m=2$ is the mass of Ps. The Ps eigenstates in the cavity are characterized by the radial and orbital quantum numbers of the center-of-mass motion, which we denote by $N$ and $L$, respectively. The means of determining $N$ and $L$ for each Ps eigenstate is detailed in Ref. \cite{Brown17}; for Ps$(1s)$ states matters are simplified by the fact that $J=L$ and $\Pi = (-1)^L$. From Eq.~(\ref{eq:E}), the corresponding Ps center-of-mass momentum $K$ is found as
\begin{equation}
K=\sqrt{4E + 1}.
\end{equation}
In Sec.~\ref{sec:pha_scs} we show how the scattering phase shifts and cross section are determined from these center-of-mass momenta.

\subsection{Scattering phase shifts and cross section}\label{sec:pha_scs}

A positive-energy electron or positron radial wave function near the cavity wall (i.e., away from the atom) has the form \cite{Burke77}
\begin{equation}\label{eq:rad}
P_{\varepsilon l}(r)\propto \sqrt{r}\left[\cos \delta _l\, J_{l+1/2}(kr)-\sin \delta _l\, Y_{l+1/2}(kr)\right],
\end{equation}
where $k=\sqrt{2\varepsilon}$, $J_{l+1/2}$ and $Y_{l+1/2}$ are the Bessel and Neumann functions, respectively, and $\delta _l$ is the phase shift. From the boundary condition $P_{\varepsilon l}(R_c)=0$, one finds the phase shift as
\begin{equation}\label{eq:del}
\tan \delta _l =\frac{J_{l+1/2}(kR_c)}{Y_{l+1/2}(kR_c)}.
\end{equation}
Similarly, one can obtain the Ps-atom scattering phase shifts $\delta_L(K)$ from the values of the Ps center-of-mass momenta $K$. Away from the atom and the cavity wall, the Ps wave function decouples into separate internal and center-of-mass parts, viz.,
\begin{equation}
\Psi_{J\Pi}(\mathbf{r},\mathbf{R}) \simeq \psi_{1s}(\vec{r}) \Phi_{1s[N,L]}(\mathbf{R}),
\end{equation}
where $\mathbf{r}=\mathbf{r}_e-\mathbf{r}_p$ and $\mathbf{R}=(\mathbf{r}_e+\mathbf{r}_p)/2$.  The phase shifts are determined from the long-range form of the center-of-mass wave function; at distances where the Ps-atom interaction is negligible, the function $\Phi_{1s[N,L]}(\mathbf{R})$ behaves as [cf. Eq.~(\ref{eq:rad})]
\begin{align}\label{eq:wf_asymp}
\Phi_{1s[N,L]}(\mathbf{R}) \propto \frac{1}{\sqrt{R}} \big[& \cos \delta_L(K)\, J_{L+1/2}(KR)\nonumber\\ 
{}-{}&\sin \delta_L(K)\, Y_{L+1/2}(KR) \big] Y_{LM_L}(\Omega_{\mathbf{R}}) ,
\end{align}
where $R=\lvert \mathbf{R} \rvert$ is the distance to the Ps center of mass. The presence of an impenetrable confining wall at $R=R_c$ is equivalent to the boundary condition
\begin{equation}\label{eq:bound_cond}
\left. \Phi_{1s[N,L]}(\mathbf{R}) \right\rvert_{R=R_c-\rho(K)} = 0,
\end{equation}
where $\rho(K)$ is the ``collisional radius'' of Ps($1s$) \cite{Brown17}. This quantity can be thought of as the distance of closest approach between the Ps center of mass and the wall. It is determined by considering states of Ps in an empty cavity, as described in Ref.~\cite{Brown17}. From Eqs.~(\ref{eq:wf_asymp}) and (\ref{eq:bound_cond}), we find the phase shift as
\begin{equation}\label{eq:phashft_form}
\tan \delta_L(K) = \frac{J_{L+1/2}(K[R_c-\rho(K)])}{Y_{L+1/2}(K[R_c-\rho(K)])}.
\end{equation}
The expression for the $S$-wave ($L=0$) phase shift is particularly simple, viz.,
\begin{equation}
\delta_0(K) = - K[R_c-\rho(K)]\quad (\operatorname{mod}\pi ).
\end{equation}
As usual, the phase shifts are determined to within $n\pi $, where $n$ is an integer. When presenting the phase shifts (see Sec.~\ref{subsec:ph}), it is convenient to require that $\delta _L\to 0$ as $K\to 0$.

The method of finding the phase shifts described above is known as the ``box-variational method.'' It has been used sporadically in various situations \cite{Percival57,Alhassid84,vanFaasen07,Cheng14}. We believe, however, that this is the first time that the method has been used to study the scattering of a composite particle.

Once the phase shifts have been found for selected values of the Ps momentum, effective-range-type fits are used to determine $\delta_L$ for an arbitrary low momentum $K$. In the frozen-target approximation, the atomic potential is of short range, and the effective-range expansion for the $L$th partial wave is \cite{Burke77}
\begin{equation}\label{eq:eff_ran_the}
\delta_L(K) \simeq \alpha_L K^{2L+1} + \beta_L K^{2L+3} + O(K^{2L+5}),
\end{equation}
where $\alpha_L$ and $\beta_L$ are constants. For small $K$ the $S$ wave dominates; its effective-range expansion is written in the form
\begin{equation}
\delta_0(K) \simeq -AK - \frac{1}{2} A^2 r_0 K^3 + O(K^5) ,
\end{equation}
where $A$ is the scattering length and $r_0$ is the effective range \cite{Burke77}.

When the van der Waals potential is included (see Sec.~\ref{subsec:vdW}), its long-range nature leads to the following modified-effective-range formulae for low $K$ \cite{Ganas72,Fabrikant14}:
\begin{subequations}
\begin{align}\label{eq:L0}
\delta_0(K) &\simeq -AK - \frac{1}{2}A^2 r_0K^3 + \zeta _0 K^4 + O(K^5) ,\\
\delta_1(K) &\simeq \alpha K^3 + \zeta _1 K^4 + \beta K^5 + \gamma K^7 \ln K + O ( K^7 ), \label{eq:L1}\\
\delta_L(K) &\simeq \zeta_L K^4 + \eta_L K^{2L+1} + \lambda_L K^{2L+3} + \mu_L K^{2L+5} \ln K \nonumber\\
&\quad{}+ O( K^{2L+5} ) \qquad (L \geq 2),\label{eq:L2}
\end{align}
\end{subequations}
where $\alpha$, $\beta$, $\gamma$, $\zeta_L$, $\eta_L$, $\lambda_L$, and $\mu_L$ are constants. The $K^4$ term in the above expansions is due to the $-C_6 / R^6$ asymptotic behavior of van der Waals interaction, and the corresponding coefficient is given by \cite{Ganas72}
\begin{equation}\label{eq:zetaL}
\zeta_L=\frac{6 m \pi C_6 }{(2L-3)(2L-1)(2L+1)(2L+3)(2L+5)},
\end{equation}
where $m=2$ is the mass of Ps and $C_6$ is the van der Waals constant. Note that $\zeta _LK^4$ is the leading term for $L\geq 2$. It leads to characteristic rise $\delta_L \propto K^4$ at low Ps momenta and enables estimates of $\delta _L$ for small $K$. 

The partial scattering cross section $\sigma_L$ for the $L$th partial wave is determined by the phase shift,
\begin{equation}\label{eq:part_cs}
\sigma_L(K) = \frac{4\pi}{K^2} (2L+1) \sin^2 \delta_L(K),
\end{equation}
with the total elastic cross section $\sigma$ given by
\begin{equation}\label{eq:cs}
\sigma(K) = \sum_{L=0}^{\infty} \sigma_L(K).
\end{equation}
In the energy range considered here ($K\leq 1$~a.u.), it is usually sufficient to include only the $S$, $P$, and $D$ waves in Eq.~(\ref{eq:cs}), though the contribution of the $F$ wave may not be negligible for heavier target atoms like Kr or Xe (see Sec.~\ref{subsec:cs2}).

For the purpose of comparison with experimental data, it is also useful to compute the momentum-transfer cross section, defined as
\begin{equation}
\sigma_m=\int (1-\cos\theta) \frac{d\sigma}{d\Omega} \, d\Omega ,
\end{equation}
where $\theta$ is the scattering angle, $d\Omega$ is the element of solid angle, and $d\sigma/d\Omega$ is the differential cross section. In terms of the phase shifts, it is given by
\begin{equation}
\sigma_m=\frac{4\pi}{K^2} \sum_{L=0}^\infty (L+1) \sin^2 \left[ \delta_{L+1}(K)-\delta_L(K) \right],
\end{equation}
and for low $K$ converges quickly with respect to the number of partial waves included. In the limit $K\to 0$ (in practice, $K\ll 1$~a.u.), the $S$ wave dominates, and the cross sections are determined by the scattering length: $\sigma_m(0)=\sigma(0)=4\pi A^2$.

\subsection{van der Waals interaction}\label{subsec:vdW}

The long-range interaction between the Ps and the target atom is described by the van der Waals potential, which arises due to mutual polarization of the atoms. At large distances it behaves as $-C_6/R^6$, where $R$ is the distance between the Ps center of mass and the nucleus of the target atom. Previous studies have shown that the van der Waals interaction may have a significant effect on Ps-atom scattering \cite{Mitroy01,Mitroy03,Fabrikant14}. 

At small $R$ the asymptotic form $-C_6/R^6$ displays an unphysical growth. Here, the electron-electron and electron-positron correlation effects that make up the van der Waals interaction cannot be described by means of a simple potential. Following Fabrikant and Gribakin \cite{Fabrikant14} and Gribakin \emph{et al.} \cite{Gribakin16}, we treat this problem by introducing an effective cutoff function in the van der Waals potential,
\begin{equation}\label{eqn:vdw_pot}
V_W(R)=-\frac{C_6}{R^6} \left\{ 1-\exp\left[ -\left( R/R_0\right)^8\right] \right\},
\end{equation}
where $R_0$ is an adjustable cutoff radius. Its magnitude is expected to be close to the sum of the atomic and Ps radii. The potential $V_W(R)$ is included by adding its matrix elements 
\begin{align}
\langle \nu' \mu' \vert V_W \vert \mu \nu \rangle &=
\iint  \varphi_{\nu'}^*(\mathbf{r}_p) \varphi_{\mu'}^*(\mathbf{r}_e) V_W \left(\lvert\mathbf{r}_e+\mathbf{r}_p\rvert /2 \right) \nonumber\\ \label{eq:m_el_VW}
&\quad{}\times \varphi_{\mu}(\mathbf{r}_e) \varphi_{\nu}(\mathbf{r}_p) \, d^3\mathbf{r}_e \, d^3\mathbf{r}_p. 
\end{align}
to the Hamiltonian matrix, Eq.~(\ref{eq:ham_mat_elem}). See Appendix~\ref{app_vdw} for details of how the matrix elements $\langle \nu' \mu' \vert V_W \vert \mu \nu \rangle$ are computed. 

Regarding the values of the van der Waals coefficients, for He we use the configuration-interaction value obtained by Mitroy and Bromley \cite{Mitroy03} ($C_6=13.37$~a.u.); for Ne, Ar, Kr, and Xe the values are taken from empirically scaled random-phase-approximation calculations of Swann \textit{et al.} \cite{Swann15} ($C_6=26.48$, $98.69$, $146.71$, and $227.38$~a.u., respectively). All of these values are expected to be accurate to within 1\%.

As the scattering cross section may be sensitive to the choice of $R_0$ (which at this stage is somewhat arbitrary), it is useful to consider the ``strength'' of this potential. The attractive strength of a generic spherically symmetric potential $U(r)$ can be estimated by the following dimensionless parameter \cite{Bargmann52,Dzuba94}:
\begin{equation}
S=2m\int_0^\infty r \vert U(r) \vert \, dr.
\end{equation}
For the potential $V_W(R)$ [Eq.~(\ref{eqn:vdw_pot})] the integration can be performed analytically, giving
\begin{equation}\label{eq:S}
S=\frac{mC_6 \sqrt{\pi}}{2R_0^4}.
\end{equation}
Thus, the strength of the van der Waals potential increases rapidly with decreasing $R_0$, confirming that the cross section may be sensitive to the choice of $R_0$. For instance, decreasing $R_0$ from a value of 3.0~a.u. to 2.5~a.u. increases the strength $S$ by a factor of 2. This sensitivity, however, is mitigated by the strongly repulsive static Ps-atom interaction, which means that the probably of finding the Ps center of mass at small $R$ is quite small.

\section{\label{sec:numimp}Numerical implementation}

First, we use a standard Hartree-Fock program \cite{Amusia97} to generate the ground-state electron orbitals of the target atom. These wave functions are used to generate the electron- and positron-atom potentials $U_e$ and $U_p$. Then, a $B$-spline basis is used to construct the single-particle electron and positron basis states $\varphi_\mu$ and $\varphi_\nu$ (see, e.g., Ref. \cite{Gribakin04}). The $B$~splines are defined on a set of radial points (or ``knots'') $r_j$ given by
\begin{equation}\label{eq:QLknots}
r_j = \begin{cases}
0 & \text{for} \quad 1 \leq j \leq k,\\
\dfrac{A(j-k)^2}{B+j-k} & \text{for} \quad k \leq j \leq n+1,\\
R_c & \text{for} \quad n+1 \leq j \leq n+k,
\end{cases}
\end{equation}
where $n$ is the number of splines used, $k$ is their order,
\begin{equation}
A=\frac{B+n+1-k}{(n+1-k)^2}R_c,
\end{equation}
and $B$~is a parameter; a value of $B=100$ has been used throughout. This knot sequence simultaneously provides an accurate representation of the atomic ground-state orbitals and the positive-energy electron and positron states \footnote{Compared with the exponential knot sequence, which is convenient for atomic calculations, including efficient convergence of many-body-theory sums \cite{Gribakin04}, the ``quadratic-linear'' knot sequence (\ref{eq:QLknots}) ensures that the Ps states, Eq. (\ref{eq:CIexpansion}), are represented accurately both at small distances and near the cavity wall.}.

In this work we carry out calculations using four different cavity radii, namely, $R_c=10$, 12, 14, and 16~a.u. Keeping the cavity radius small assists convergence of the expansion (\ref{eq:CIexpansion}). For $D$-wave scattering with the van der Waals interaction, the phase shifts are very small (see Fig.~\ref{fig:phase_shifts_vdw}), and a fair amount of scatter was observed among the data points. We therefore ran the $D$-wave calculations using an additional value, $R_c=15$~a.u., to reduce the uncertainty in the effective-range fits. In any case, the $D$-wave partial elastic cross sections are small (in the van der Waals approximation) and contribute little to the elastic cross section in the energy range considered. A set of 60 $B$~splines of order 9 has been used throughout.

To study $S$-, $P$-, and $D$-wave scattering of Ps($1s$), it is necessary to construct Ps states in the field of each of the five target atoms with $J^\Pi=0^+$, $1^-$, and $2^+$ respectively. The dimension $\mathcal{N}$ of the Hamiltonian matrix is 
\begin{equation}\label{eq:dimH}
\mathcal{N}=\begin{cases}
n_\text{max}^2 (l_\text{max}+1) & \text{for} \quad J^\Pi=0^+ , \\
2 n_\text{max}^2 l_\text{max} & \text{for} \quad J^\Pi=1^-, \\
n_\text{max}^2 (3 l_\text{max}-2) & \text{for} \quad J^\Pi=2^+,
\end{cases}
\end{equation}
where $l_\text{max}$ is the maximum orbital angular momentum and $n_\text{max}$ is the number of radial states for each $l\leq l_\text{max}$, in the electron and positron basis sets. To keep the size of the calculation manageable, we used $l_\text{max}=n_\text{max}=20$ for $J^\Pi=0^+$, $l_\text{max}=n_\text{max}=18$ for $J^\Pi=1^-$, and $l_\text{max}=n_\text{max}=16$ for $J^\Pi=2^+$. Once the Hamiltonian matrix (\ref{eq:ham_mat_elem}) has been diagonalized and the Ps energy levels found, we identify those which correspond to $1s$ states of Ps and have $K<1$~a.u. Their energies are extrapolated to the limits $l_\text{max}\to\infty$ and $n_\text{max}\to\infty$; see Ref. \cite{Brown17} for full details of the procedure. The scattering phase shifts and cross sections are then found as described in Sec.~\ref{sec:theory}.

Note that in Ref. \cite{Brown17} the Ps($1s$) collisional radius $\rho(K)$ was given as a linear fit in terms of $K$. In the present work we use more accurate fits for the calculated data in Ref. \cite{Brown17} and account for the fact that the radius $R_c$ of the cavity may have an effect on $\rho$. See Appendix~\ref{sec:coll_rad} for the fits used here.

\section{Results}\label{sec:res}

\subsection{Scattering phase shifts}\label{subsec:ph}

\begin{figure*}
\includegraphics{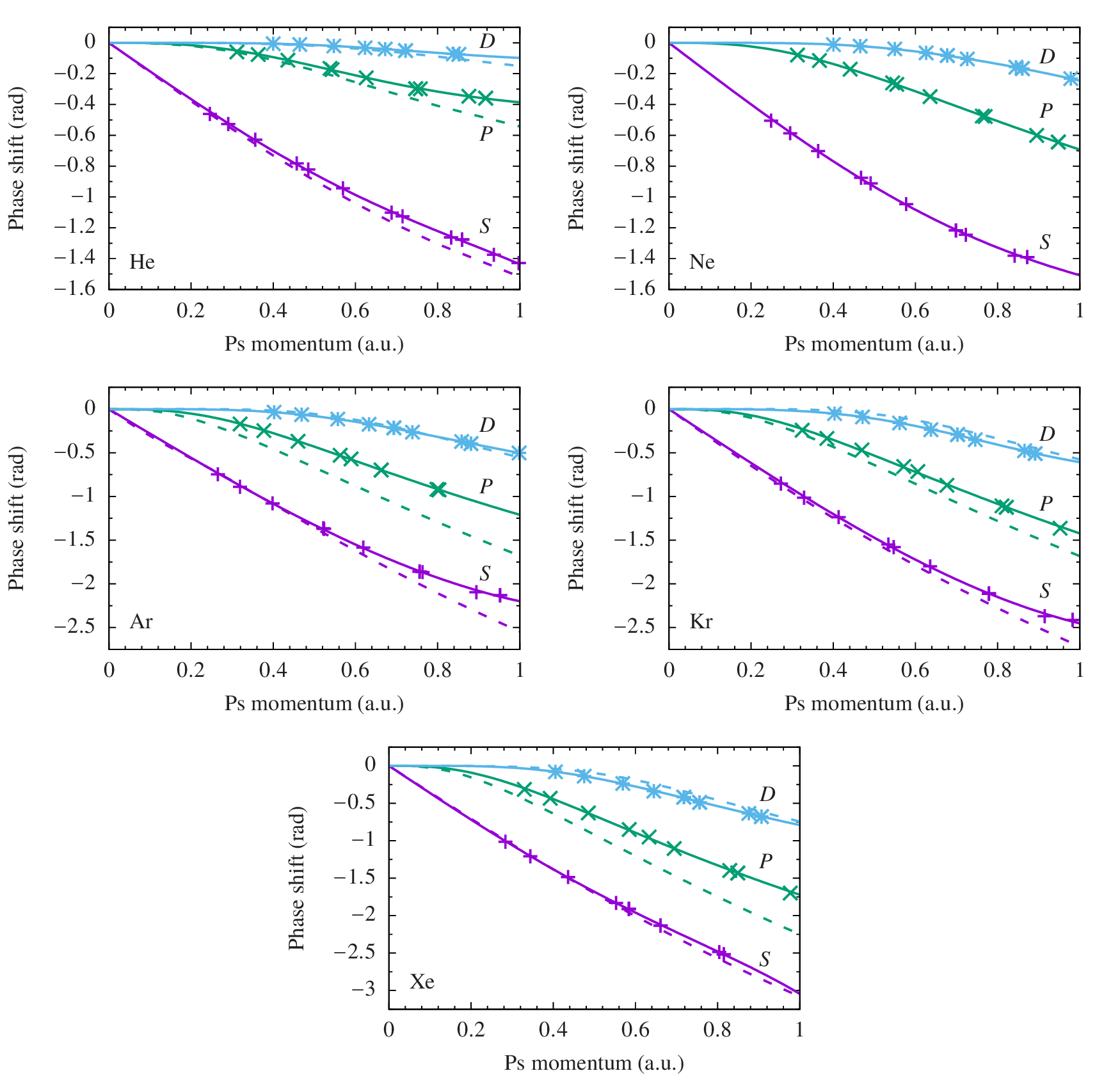}
\caption{\label{fig:phase_shifts}$S$-, $P$-, and $D$-wave scattering phase shifts for Ps scattering on He, Ne, Ar, Kr, and Xe in the frozen-target approximation. Purple plusses, calculated $\delta_0$; green crosses, calculated $\delta_1$; blue asterisks, calculated $\delta_2$; solid purple line, effective-range fit for $\delta_0$; solid green line, effective-range fit for $\delta_1$; solid blue line, effective-range fit for $\delta_2$. The dashed lines show results of existing calculations in the same colors as the present results; for He these are the static-exchange phase shifts of Sarkar and Ghosh \cite{Sarkar97}, for Ar and Kr they are the pseudopotential phase shifts of Fabrikant and Gribakin \cite{Fabrikant14}, and for Xe they are the pseudopotential phase shifts of Gribakin \textit{et al.} \cite{Gribakin16}.}
\end{figure*}

Figure~\ref{fig:phase_shifts} shows the $S$-, $P$-, and $D$-wave phase shifts for Ps scattering on He, Ne, Ar, Kr, and Xe in the frozen-target approximation. Shown by crosses are the phase shifts obtained from Eq.~(\ref{eq:phashft_form}) for different center-of-mass motion states of Ps($1s$) and different cavity radii $R_c$. Solid lines represent effective-range-type fits of the data (see Appendix~\ref{sec:tables} for details). The phase shifts are negative, which indicates that in the frozen-target approximation the Ps-atom interaction is repulsive.
Also shown are the phase shifts from the static-exchange calculations for He \cite{Sarkar97} and pseudopotential calculations for Ar, Kr, and Xe \cite{Fabrikant14,Gribakin16}.

Compared to the frozen-target approximation, the static-exchange calculations do not allow for the distortion of Ps in the field of the atom. Nevertheless, there is close agreement between the two sets of data for He. At increasing Ps momenta, the static-exchange phase shifts \cite{Sarkar97} are slightly more negative than the present phase shifts. This could be expected, since the distortion of Ps allowed by the frozen-target approximation softens the Ps-atom repulsion. A similar difference between the static-exchange and frozen-target approximation was noted earlier by Blackwood \textit{et al.} for He, Ne and Ar \cite{Blackwood99,Blackwood02}.

For Ar, Kr, and Xe, there is fairly good agreement with the pseudopotential calculations \cite{Fabrikant14,Gribakin16}, in which the pseudopotential was designed to represent the Ps-atom interaction at the static-exchange level. The largest discrepancy in each case is for the $P$ wave. As noted by Gribakin \textit{et al.} \cite{Gribakin16}, the pseudopotential method likely overestimates the $P$-wave contribution to the scattering, so the present $P$-wave phase shifts for Ar, Kr, and Xe should be considered superior at this level of approximation.

\begin{figure*}
\includegraphics{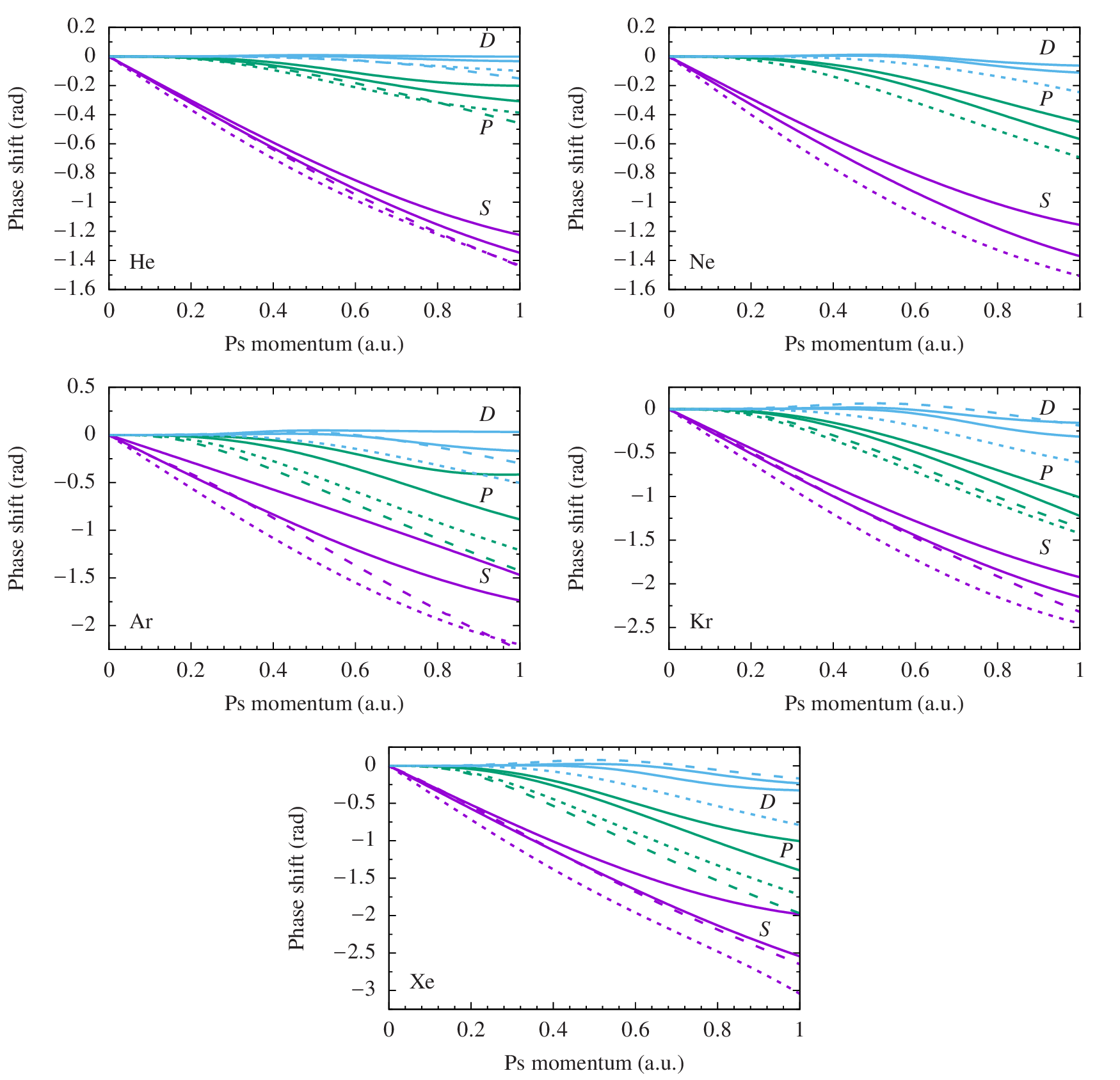}
\caption{\label{fig:phase_shifts_vdw}$S$-, $P$-, and $D$-wave scattering phase shifts for Ps scattering on He, Ne, Ar, Kr, and Xe with inclusion of the van der Waals interaction. Solid purple lines, effective-range fits for $\delta_0$; solid green lines, effective-range fits for $\delta_1$; solid blue lines, effective-range fits for $\delta_2$. For each partial wave, the lower (higher) curve corresponds to the larger (smaller) cutoff radius (see text for values of $R_0$). Dotted lines in the same colors show the fits of the frozen-target phase shifts presented in Fig.~\ref{fig:phase_shifts}.
Dashed lines in the same colors show the results of other calculations that include the van der Waals interaction: for He these are the static-exchange--with--van-der-Waals phase shifts of Barker and Bransden \cite{Barker69}, and the pseudopotential phase shifts of Fabrikant and Gribakin \cite{Fabrikant14} (Ar and Kr) and Gribakin \textit{et al.} \cite{Gribakin16} (Xe).}
\end{figure*}

Figure~\ref{fig:phase_shifts_vdw} shows the effect of including the van der Waals interaction on the phase shifts. Here we only present the effective-range-type fits of the data (see Appendix~\ref{sec:tables} for the fits used), to avoid cluttering the plots.
To examine the sensitivity of the results to the choice of the cutoff radius, 
we have used two values for each target atom:  $R_0=2.5$~a.u. and 3.0~a.u. for He, Ne, and Ar; $R_0=3.0$~a.u. and 3.5~a.u. for Kr and Xe.
These values are similar to those used in Refs. \cite{Fabrikant14,Gribakin16}, which ensured that the calculated cross section merged with experimental data for $K>1$~a.u.
Since the van der Waals interaction is attractive, the corresponding phase shifts lie higher than their frozen-target counterparts, and the $D$-wave phase shifts are positive at small $K$, as predicted by Eqs.~(\ref{eq:L2}) and (\ref{eq:zetaL}). In spite of the strong dependence of the van der Waals strength parameter (\ref{eq:S}) on the cutoff radius, the phase shifts obtained using two different values of $R_0$ for each atom are quite close. 
This is a consequence of the strong static Ps-atom repulsion, which reduces the probability of finding the Ps center of mass at small distances, where the short-range part of the model van der Waals potential (\ref{eqn:vdw_pot}) is strongest.

Also shown in Fig.~\ref{fig:phase_shifts_vdw} are static-exchange--with--van-der-Waals phase shifts of Barker and Bransden \cite{Barker69} for He, and pseudopotential calculations with the van der Waals interaction for Ar, Kr, and Xe \cite{Fabrikant14,Gribakin16}. There is reasonably good agreement with these calculations, except for $S$-wave scattering on Ar and $P$-wave scattering on Ar and Xe.

\subsection{Scattering lengths and cross sections}\label{subsec:cs}

We will consider scattering lengths and elastic cross sections for He separately from the other four atoms.

\subsubsection{Ps scattering on He}\label{subsec:cs1}

For Ps-He scattering a scattering length of $A=1.86$~a.u. was obtained in the frozen-target approximation, corresponding to a zero-energy cross section of $13.8 \pi a_0^2$. With inclusion of the van der Waals potential this reduced to $A=1.61$~a.u. for $R_0=3.0$~a.u., or $A=1.52$~a.u. for $R_0=2.5$~a.u., giving zero-energy cross sections of $10.4\pi a_0^2$ and $9.2\pi a_0^2$, respectively. Table~\ref{tab:He_data} shows these data along with a selection of previous results.
\begin{table}
\caption{\label{tab:He_data}Scattering lengths $A$ (in units of $a_0$) and zero-energy cross sections $\sigma(0)$ (in units of $\pi a_0^2$) for Ps-He scattering.}
\begin{ruledtabular}
\begin{tabular}{lcc}
Method & $A$ & $\sigma(0)$ \\
\hline
Present, frozen target & 1.86 & 13.8 \\
Present, van der Waals, $R_0=3.0$~a.u. & 1.61 & 10.4 \\
Present, van der Waals, $R_0=2.5$~a.u. & 1.52 & 9.2 \\
Static exchange (not converged) \cite{Fraser66} & 2.17 & 18.8 \\
Static exchange \cite{Fraser68} & 1.88 & 14.2 \\
Static exchange \cite{Barker68} & 1.80 & 13.0 \\
Static exchange with van der Waals \cite{Barker69} & 1.61 & 10.4 \\
Kohn variational, static exchange \cite{Drachman70} & 1.72 & 11.8 \\
$R$ matrix, static (no exchange) \cite{McAlinden96} & & {>}20 \\
$R$ matrix, static exchange \cite{Blackwood99} & 1.91 & 14.6 \\
$R$ matrix, 22 Ps states \cite{Blackwood99} & 1.82 & 13.2 \\
$T$ matrix, model static exchange \cite{Biswas00} & 1.03 & 4.2 \\
$T$ matrix, 3 Ps states, model exchange \cite{Adhikari00} & 0.91 & 3.3 \\
$T$ matrix, 1 Ps state, 3 He states \cite{Ghosh01} & 1.39 & 7.7 \\
$T$ matrix, 2 Ps states, 3 He states \cite{Ghosh01} & 1.36 & 7.4 \\
$T$ matrix, static exchange \cite{Basu01} & 1.93 & 14.9 \\
$T$ matrix, 3 Ps states \cite{Basu01} & 1.92 & 14.7 \\
$T$ matrix, 2 Ps states, 3 He states \cite{Basu01} & 1.36 & 7.4 \\
Stochastic variational, frozen target \cite{Mitroy01} & 1.84 & 13.5 \\
Stochastic variational, van der Waals \cite{Mitroy01} & 1.57 & 9.8 \\
Diffusion Monte Carlo \cite{Chiesa02} & 1.40 & 7.9 \\
Kohn variational, 3 Ps states \cite{DiRenzi03} & 1.60 & 10.2 \\
$R$ matrix, 9 Ps states, 9 He states \cite{Walters04} & 1.6 & 9.9 \\
Recommended & $1.60\pm 0.1$ & $10.2\pm 1.3$
\end{tabular}
\end{ruledtabular}
\end{table}
%
%
%

Blackwood \textit{et al.} \cite{Blackwood99} used the $R$-matrix method in a variety of approximations. Their static-exchange calculation gave $A=1.91$~a.u., in contrast with the earlier converged result of $A=1.80$~a.u. of Barker and Bransden \cite{Barker68}. A contributing factor to the discrepancy between these values may be that Blackwood \textit{et al.} \cite{Blackwood99} used a Hartree-Fock wave function for the He atom, whereas Barker and Bransden \cite{Barker68} used only a single Slater-type orbital. Blackwood \textit{et al.} \cite{Blackwood99} then allowed for distortion of the Ps projectile using a channel space of 22 coupled pseudostates, but retained the He atom in its ground state. The scattering length of this calculation was $A=1.82$~a.u., in agreement with the present frozen-target result at the level of 2\%.

Basu \textit{et al.} \cite{Basu01} used the momentum-space $T$-matrix approach in the static-exchange approximation, obtaining $A=1.93$~a.u., in good agreement with the $R$-matrix value \cite{Blackwood99}. They then allowed for limited distortion of the Ps atom by performing a calculation with the Ps($1s$), Ps($2s$), and Ps($2p$) states, and the scattering length changed by only 0.5\%. They also considered excitations of the target He atom into its $1s \, 2s \, ^1S^e$ and $1s \, 2p \, ^1P^o$ states. The inclusion of these excitations led to a greatly reduced scattering length of $1.36$~a.u. (using only the Ps($1s$) and Ps($2p$) states). Similar results were obtained by Ghosh \textit{et al.} \cite{Ghosh01}. This indicates that excitations of the target atom (which can give rise to the van der Waals interaction) are important and should be included. However, their values are about 10\% lower than the smaller of the present two van der Waals scattering lengths (1.52~a.u. for $R_0=2.5$~a.u.).

There were also $T$-matrix calculations by Biswas and Adhikari \cite{Biswas00} and Adhikari \cite{Adhikari00}, but they are in serious conflict with those of Ghosh and coworkers \cite{Ghosh01,Basu01}, who used an identical method. Mitroy and Ivanov \cite{Mitroy01} stated that Adhikari and coworkers' model exchange interaction was of ``dubious validity'' and noted that doubt had already been raised about these calculations by other authors \cite{Blackwood99,Ghosh01a}.

Mitroy and Ivanov \cite{Mitroy01} employed the fixed-core stochastic-variational method, allowing for distortion of the Ps projectile but not the He atom, and found $A=1.84$~a.u., which is within 1\% of the present frozen-target value. Model electron- and positron-atom polarization potentials were then added, leading to a scattering length of $1.57$~a.u. (a 15\% reduction), which is within 2--3\% of our van der Waals values. These data are in close agreement with the earlier static-exchange \cite{Barker68} and static-exchange--with--van-der-Waals \cite{Barker69} calculations. We note again the importance of accounting for distortion of the target atom. The most sophisticated $R$-matrix calculation of Walters \textit{et al.} \cite{Walters04} included nine Ps states and nine He states, leading to a scattering length of $1.6$~a.u., supporting the results of the earlier van der Waals calculations \cite{Barker69,Mitroy01} and in near-perfect agreement with the present van der Waals value of 1.61~a.u. (for $R_0=3.0$~a.u.). Allowing for a ${\sim}5$\% error in the scattering length, we show recommended values of the scattering length and zero-energy cross section in the last line of Table~\ref{tab:He_data}.

\begin{figure}
\includegraphics{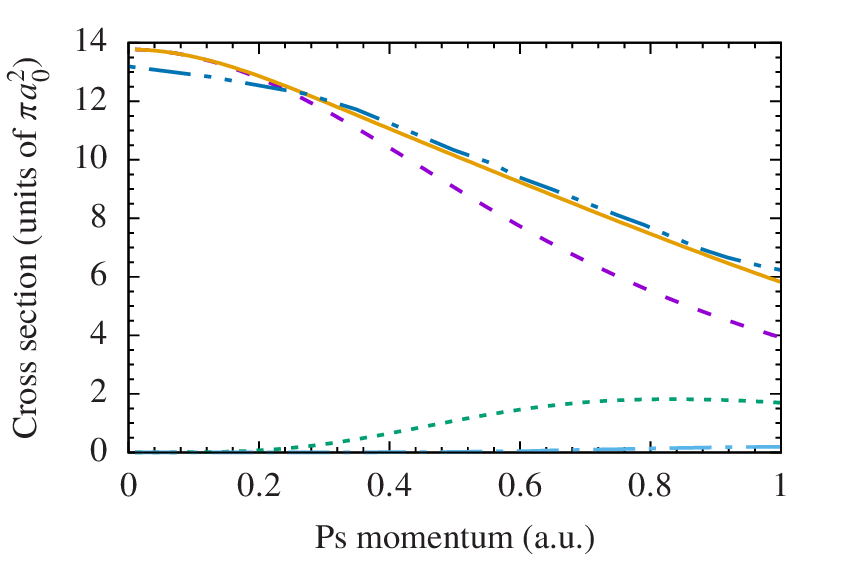}
\caption{\label{fig:He_cs}Partial and total elastic cross sections for Ps-He scattering in the frozen-target approximation. Dashed purple line, $\sigma_0$; dotted green line, $\sigma_1$; dash-dotted cyan line, $\sigma_2$; solid orange line, $\sigma=\sigma_0+\sigma_1+\sigma_2$; dash-double-dotted navy line, 22-Ps-state calculation of $\sigma$ of Blackwood \textit{et al.} \cite{Blackwood99}.}
\end{figure}

Figure~\ref{fig:He_cs} shows the partial and total elastic cross sections for Ps-He scattering in the frozen-target approximation. 
One can see that the $P$ wave is unimportant for $K \lesssim 0.2$~a.u. and the $D$ wave is unimportant for $K \lesssim 0.6$~a.u. Even at $K=1$~a.u. the $D$ wave adds little to the total elastic cross section. Also shown on the graph is the result of the $R$-matrix calculation of Blackwood \textit{et al.} \cite{Blackwood99}, where 22 Ps states were included but the He atom was frozen. The small difference in the region $K\leq0.2$~a.u., may be related, at least in part, to uncertainties of our fit of the $S$-wave phase shift at low $K$ (cf.~Fig.\ref{fig:phase_shifts}). This difference aside, the agreement between the cross sections obtained in the same approximation using two very different methods confirms the validity of the present method of solving the Ps-atom scattering problem using a hard-wall cavity.

\begin{figure}
\includegraphics{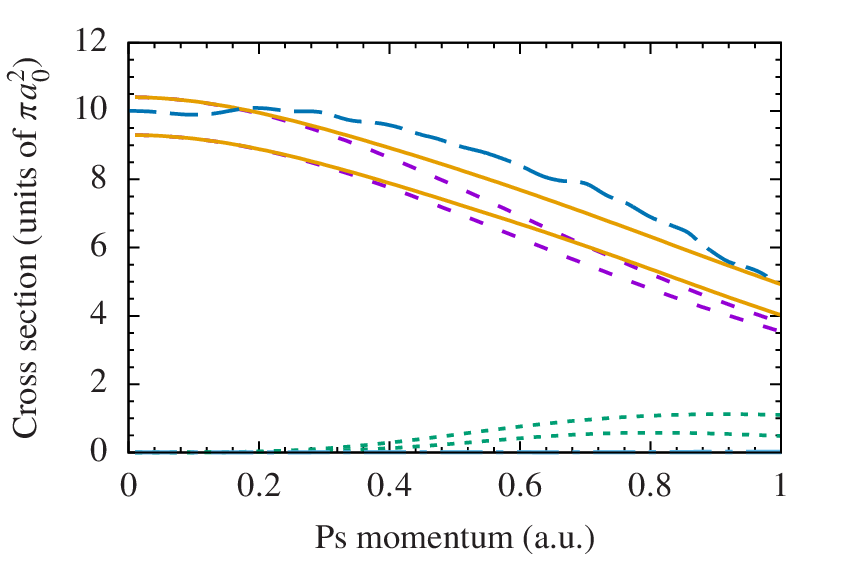}
\caption{\label{fig:He_cs_vdw}Partial and total elastic cross sections for Ps-He scattering with inclusion of the van der Waals interaction. Dashed purple lines, $\sigma_0$; dotted green lines, $\sigma_1$; dash-dotted cyan lines, $\sigma_2$; solid orange lines, $\sigma=\sigma_0+\sigma_1+\sigma_2$. For each cross section, the higher curve corresponds to $R_0=3.0$~a.u.,
the lower  to $R_0=2.5$~a.u. Long-dashed navy line, 9-Ps-9-He-state calculation of $\sigma$ of Walters \textit{et al.} \cite{Walters04}.}
\end{figure}

Figure~\ref{fig:He_cs_vdw} shows the partial and total elastic cross sections for Ps-He scattering with inclusion of the van der Waals interaction. 
Adding the attractive van der Waals potential to the repulsive frozen-target Ps-atom interaction leads to a 20--30\% reduction in the cross section. The cross sections for $R_0=2.5$~a.u. and 3.0~a.u. are quite close; there is only a 13\% difference at zero energy. This relative insensitivity to the value of the cutoff radius may seem surprising, given that the minimum of $V_W(R)$, Eq. (\ref{eqn:vdw_pot}), changes from $-0.012$~a.u. at $R=2.8$~a.u. (for $R_0=3.0$~a.u.) to $-0.036$~a.u. at $R=2.3$~a.u. (for $R_0=2.5$~a.u.). However, for $R<3$~a.u., the relative effect of static repulsion is already quite strong, so that the change of $V_W(R)$ has a relatively small effect.
The $D$-wave partial cross section is negligible across the energy range considered and barely visible in Fig.~\ref{fig:He_cs_vdw}. Also shown in Fig.~\ref{fig:He_cs_vdw} is the 9-Ps-9-He-state calculation of Walters \textit{et al.} \cite{Walters04}, which accounted for Ps and target excitation rigorously using pseudostates. Their result is close to our cross section for $R_0=3.0$~a.u. This shows that a model-potential approach with an adjustable cutoff can be used to account for the Ps-atom dispersion forces.

\begin{figure}
\includegraphics{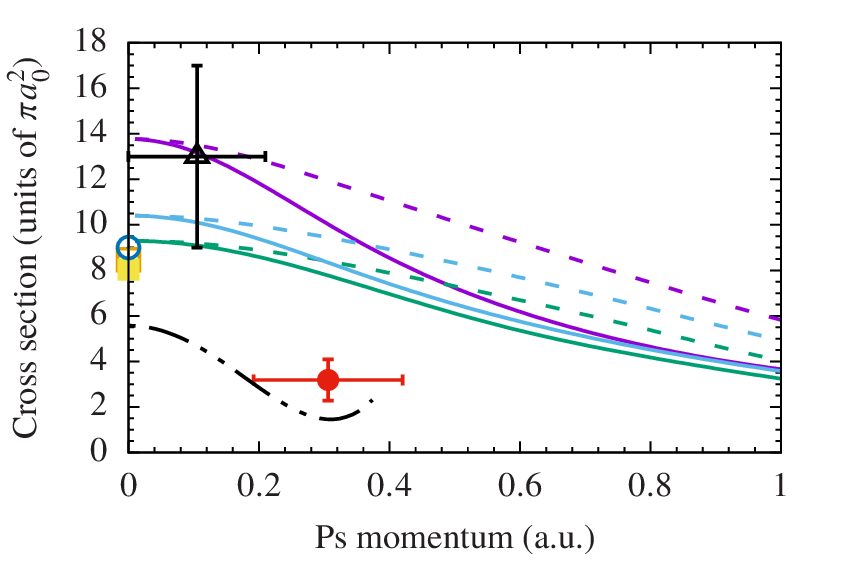}
\caption{\label{fig:He_MTCS}Momentum-transfer and elastic cross sections for Ps-He scattering. In the order of decreasing magnitude: solid (dashed) purple line, momentum-transfer (elastic) cross section in the frozen-target approximation; solid (dashed) cyan line, momentum-transfer (elastic) cross section for $R_0=3.0$~a.u.; solid (dashed) green line, momentum-transfer (elastic) cross section for $R_0=2.5$~a.u.  Measured momentum-transfer cross section: open orange square, Canter \textit{et al.} \cite{Canter75}; filled yellow square, Rytsola \textit{et al.} \cite{Rytsola84}; open navy circle, Coleman \textit{et al.} \cite{Coleman94}; filled red circle, Skalsey \textit{et al.} \cite{Skalsey03}; open black triangle, Nagashima \textit{et al.} \cite{Nagashima98}; dash-double-dotted black line, Engbrecht \textit{et al.} \cite{Engbrecht08}.}
\end{figure}

Figure~\ref{fig:He_MTCS} shows the calculated momentum-transfer cross sections along with some experimental data; also shown are our elastic cross sections. 
Away from $K=0$, the momentum-transfer cross section drops below the elastic cross section, rapidly in the frozen-target approximation, but more gradually when the van der Waals interaction is included. This is due to the interference between the $S$ and $P$ waves, and due to noticeable decrease in the absolute values of the $P$-wave phase shifts produced by the van der Waals interaction. At higher energies, the three momentum-transfer cross sections (frozen target, and van der Waals with $R_0=3.0$~a.u. and 2.5~a.u.) coalesce much sooner than the analogous elastic cross sections. The zero-energy cross sections of Canter \textit{et al.} \cite{Canter75}, Rytsola \textit{et al.} \cite{Rytsola84}, and Coleman \textit{et al.} \cite{Coleman94} are in good agreement with our $R_0=2.5$~a.u. cross section. The measurement by Nagashima \textit{et al.} \cite{Nagashima98} is in perfect agreement with our frozen-target calculation, and owing to the large error bars, is compatible with the van der Waals cross sections. Our calculations do not agree with the measurements of 
Skalsey \textit{et al.} \cite{Skalsey03} or Engbrecht \textit{et al.} \cite{Engbrecht08}, though we note that the results of Engbrecht \textit{et al.} \cite{Engbrecht08} disagree with the other low-energy measurements.

\subsubsection{Ps scattering on Ne, Ar, Kr, and Xe}\label{subsec:cs2}

Table~\ref{tab:NeArKrXe_data} shows the scattering lengths and zero-energy cross sections obtained for Ps scattering on Ne, Ar, Kr, and Xe from the present and earlier calculations.
\begin{table}
\caption{\label{tab:NeArKrXe_data}Scattering lengths $A$ (in units of $a_0$) and zero-energy cross sections $\sigma(0)$ (in units of $\pi a_0^2$) for Ps scattering on Ne, Ar, Kr, and Xe.}
\begin{ruledtabular}
\begin{tabular}{lcc}
 Method & $A$ & $\sigma(0)$ \\
\hline
\multicolumn{3}{c}{Ps-Ne calculations} \\
 Present, frozen target & 2.02 & 16.4 \\
 Present, van der Waals, $R_0=3.0$~a.u. & 1.66 & 11.0 \\
 Present, van der Waals, $R_0=2.5$~a.u. & 1.46 & 8.5 \\
 $T$ matrix, model static exchange \cite{Biswas00} & 1.41 & 8.0 \\
 Stochastic variational, frozen target \cite{Mitroy01} & 2.02 & 16.1 \\
 Stochastic variational, van der Waals \cite{Mitroy01} & 1.55 & 9.6 \\
 $R$ matrix, 22 Ps states \cite{Blackwood02} & 2.0 & 16.4 \\
 Recommended & $1.65\pm 0.15$ & $10.9\pm 2$ \\
 \multicolumn{3}{c}{Ps-Ar calculations} \\
  Present, frozen target & 2.81 & 31.6 \\
  Present, van der Waals, $R_0=3.0$~a.u. & 2.16 & 18.7 \\
  Present, van der Waals, $R_0=2.5$~a.u. & 1.43 & 8.2 \\
 $T$ matrix, model static exchange \cite{Biswas00} & 1.65 & 10.9 \\
 Stochastic variational, frozen target \cite{Mitroy01} & 2.85 & 32.8 \\
 Stochastic variational, van der Waals \cite{Mitroy01} & 1.79 & 12.8 \\
 $R$ matrix, 22 Ps states \cite{Blackwood02} & 2.8 & 32.3 \\
 Pseudopotential, static exchange \cite{Fabrikant14} & 3.19 & 40.7 \\
 Pseudopotential, van der Waals \cite{Fabrikant14} & 2.14 & 18.3 \\
 Recommended & $2.0\pm 0.2$ & $16\pm 3$ \\
  \multicolumn{3}{c}{Ps-Kr calculations} \\
  Present, frozen target & 3.11 & 38.7 \\
  Present, van der Waals, $R_0=3.5$~a.u. & 2.56 & 26.2 \\
  Present, van der Waals, $R_0=3.0$~a.u. & 2.26 & 20.4 \\
 Stochastic variational, frozen target \cite{Mitroy03} & 3.18 & 40.5  \\
 Stochastic variational, van der Waals \cite{Mitroy03} & 1.98 & 15.6 \\
  $R$ matrix, static exchange \cite{Blackwood02} & 3.3 & 44 \\
 Pseudopotential, static exchange \cite{Fabrikant14} & 3.32 & 44.1 \\
 Pseudopotential, van der Waals \cite{Fabrikant14} & 2.35 & 22.1 \\
 Recommended & $2.3\pm 0.3$ & $21\pm 6$ \\
   \multicolumn{3}{c}{Ps-Xe calculations} \\
  Present, frozen target & 3.65 & 53.3 \\
  Present, van der Waals, $R_0=3.5$~a.u. & 2.88 & 33.2 \\
  Present, van der Waals, $R_0=3.0$~a.u. & 2.63 & 27.7 \\
 Stochastic variational, frozen target \cite{Mitroy03} & 3.82 & 58.5  \\
 Stochastic variational, van der Waals \cite{Mitroy03} & 2.29 & 20.9 \\
 $R$ matrix, static exchange \cite{Blackwood03} & 3.77 & 56.9 \\
 Pseudopotential, static exchange \cite{Gribakin16} & 3.57 & 51.0 \\
 Pseudopotential, van der Waals \cite{Gribakin16} & 2.45 & 24.0 \\
 Recommended & $2.6\pm 0.3$ & $27\pm 7$ \\
\end{tabular}
\end{ruledtabular}
\end{table}
We note again that the $T$-matrix calculations \cite{Biswas00} for Ne and Ar are of questionable reliability; hence, we do not discuss them below.

We first consider the frozen-target results. For Ps scattering on Ne, we obtained a scattering length of $2.02$~a.u., in exact agreement with the frozen-target stochastic-variational calculation of Mitroy and Ivanov \cite{Mitroy01}, and in close agreement with the 22-Ps-state close-coupling calculation \cite{Blackwood02}. For Ps scattering on Ar, Kr, and Xe, the agreement with Mitroy and coworkers \cite{Mitroy01,Mitroy03} is at the level of 1.4\%, 2.2\%, and 4.5\% respectively. 
For Ar, Kr, and Xe we can also compare with the static-exchange pseudopotential results of Fabrikant and Gribakin \cite{Fabrikant14} and Gribakin \textit{et al.} \cite{Gribakin16}; the respective relative differences are 12\%, 6.3\%, and 2.2\%. Bearing in mind that the pseudopotential method uses only approximate electron- and positron-atom potentials and does not allow for distortion of the Ps projectile, the present scattering lengths should be considered superior. However, the agreement between the present method and the pseudopotential method increases as we move down the noble-gas group of the Periodic Table. This may imply that the pseudopotential method works best for heavier atoms. The 22-Ps-state close-coupling calculation \cite{Blackwood02} for Ar is also in excellent agreement with ours (within 1\%).

We now turn to the calculations that include the van der Waals interaction. 
As in He, the van der Waals potential mitigates the Ps-atom repulsion of the frozen-target approximation and reduces the magnitudes of the scattering lengths.
For Ne we obtained $A=1.66$~a.u. (1.43~a.u.) using a cutoff radius of $R_0=3.0$~a.u. (2.5~a.u.); this is in agreement with the result of Mitroy and Ivanov \cite{Mitroy01} at the level of 7\% (6\%). 
Similarly to He, the results obtained with the two cutoff radii are quite close.
For Ar, the $C_6$ constant is almost four times larger than for Ne, and the effect of the van der Waals potential is considerably greater. Here, the two scattering lengths we obtained are 2.16~a.u. ($R_0=3.0$~a.u.) and 1.43~a.u. ($R_0=2.5$~a.u.), in agreement with the results of Mitroy and Ivanov \cite{Mitroy01} at the level of $\sim$20\%. The value for the larger cutoff ($2.16$~a.u.) is close to the pseudopotential value of 2.14~a.u. \cite{Fabrikant14}. 
Moving to Kr, for which we have used $R_0=3.5$ and 3.0~a.u., our scattering lengths are respectively 30\% and 14\% greater than the stochastic-variational calculation \cite{Mitroy03}; however, the agreement with the pseudopotential calculation \cite{Fabrikant14} is much better, with an error of 4--9\%.
Lastly, for Xe, the present scattering lengths are 15--25\% larger than the stochastic-variational value \cite{Mitroy03} and 7--18\% larger than the pseudopotential value \cite{Gribakin16}.
It is worth noting that for both Kr and Xe, the scattering lengths we obtained for $R_0=3.0$ and 3.5~a.u. differ by only 10--15\%.

One reason for the discrepancies with the stochastic-variational calculations \cite{Mitroy01,Mitroy03} may be that they account for the long-range Ps-atom interactions by means of electron- and positron-atom polarization potentials along with the two-body polarization potential [see Eqs.~(8)--(10) in Ref. \cite{Mitroy01}]. This does lead to an effective van der Waals potential $-C_6/R^6$ at large $R$, with the van der Waals constant given by $C_6=\alpha \langle \rho^2 \rangle$, where $\alpha$ is the static dipole polarizability of the target atom and $\langle \rho^2 \rangle=12$~a.u. is the mean squared radius of ground-state Ps \cite{Fabrikant14}. For example, estimating $C_6$ for Ps-Xe interactions in this way gives $C_6\approx336$~a.u. (taking $\alpha\approx28$~a.u. \cite{onlinepolarizabilities}), when the true value is $C_6\approx 227$~a.u. \cite{Swann15}. It therefore appears that Mitroy and coworkers \cite{Mitroy01,Mitroy03} may have overestimated the strength of the long-range Ps-atom interaction, leading to smaller scattering lengths.

Table~\ref{tab:NeArKrXe_data} also lists our recommended values of the scattering lengths and zero-energy cross sections, based on the available data. Their uncertainties are related to the choice of the cutoff radius and the discrepancies between different methods.

\begin{figure*}
\includegraphics{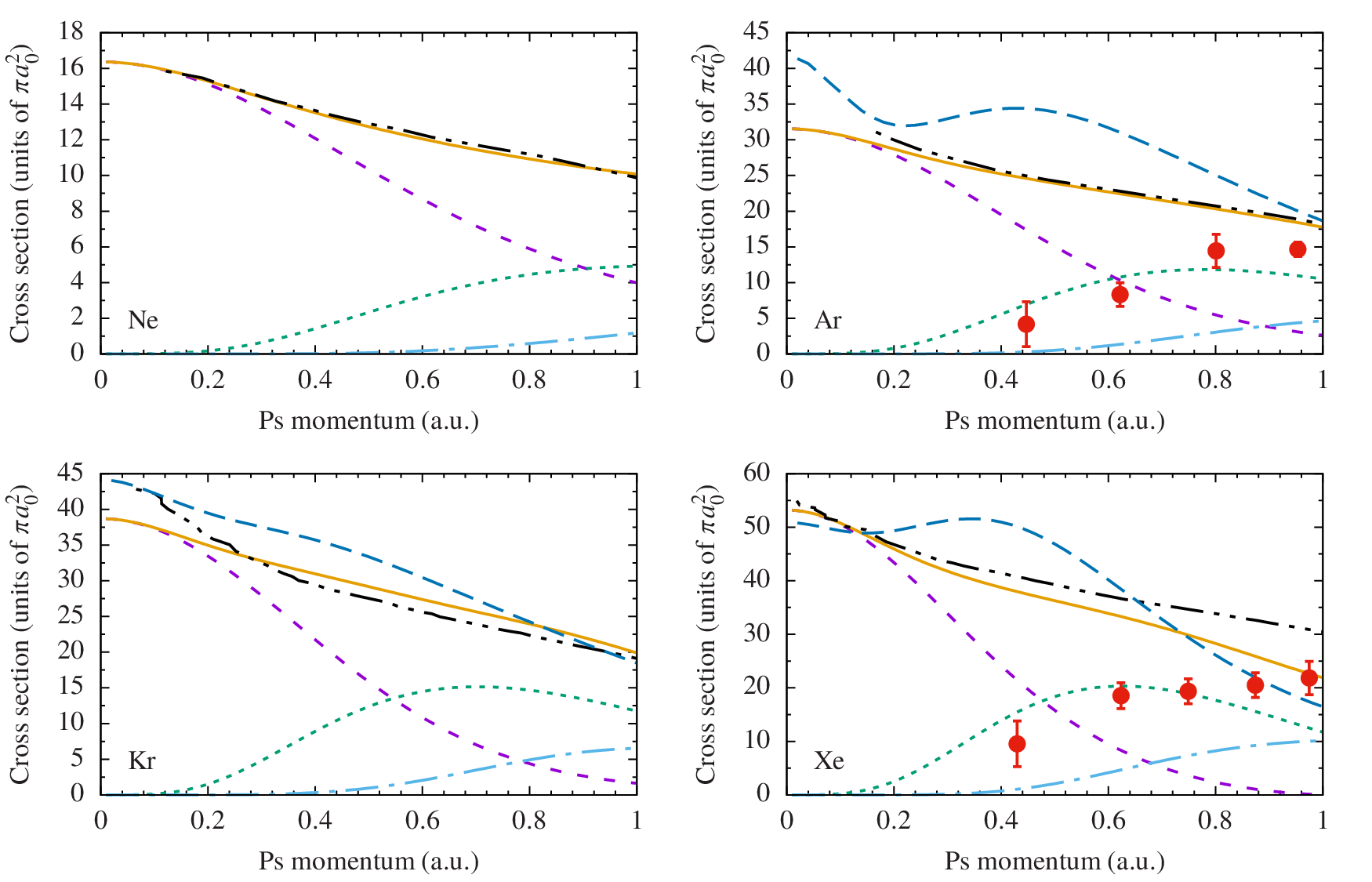}
\caption{\label{fig:NeArKrXe_cs}Partial and total elastic cross sections for Ps scattering on Ne, Ar, Kr, and Xe in the frozen-target approximation. Short-dashed purple line, $\sigma_0$; dotted green line, $\sigma_1$; dash-dotted blue line, $\sigma_2$; solid orange line, $\sigma=\sigma_0+\sigma_1+\sigma_2$; dash-double-dotted black line, calculation of $\sigma$ of Blackwood \textit{et al.} \cite{Blackwood02,Blackwood03} (using 22 Ps states for Ne and Ar but only ground-state Ps for Kr and Xe); long-dashed navy line, static-exchange pseudopotential calculation of $\sigma$ \cite{Fabrikant14,Gribakin16}; filled red circles, experimental data \cite{Brawley15}.}
\end{figure*}

Figure~\ref{fig:NeArKrXe_cs} shows the partial and total cross sections for Ps scattering on Ne, Ar, Kr, and Xe in the frozen-target approximation.
The contribution of the $P$ wave is negligible for $K<0.2$~a.u., and the $D$ wave only gives a noticeable contribution for $K \gtrsim 0.5$~a.u. Agreement of the elastic cross section for Ne and Ar with the 22-Ps-state calculations of Blackwood \textit{et al.} \cite{Blackwood02} is excellent. For Kr and Xe, Blackwood \textit{et al.} \cite{Blackwood02,Blackwood03} only performed static-exchange calculations; the agreement is still generally good, though there are some discrepancies at low energies for Kr and high energies for Xe. The pseudopotential calculations \cite{Fabrikant14,Gribakin16} should be equivalent to the static-exchange calculation. Compared to the other calculations, they show too much structure for Ar and Xe, which may be due to an overestimated $P$-wave contribution (cf. Fig.~\ref{fig:phase_shifts}), but look reasonable for Kr.

\begin{figure}
\includegraphics{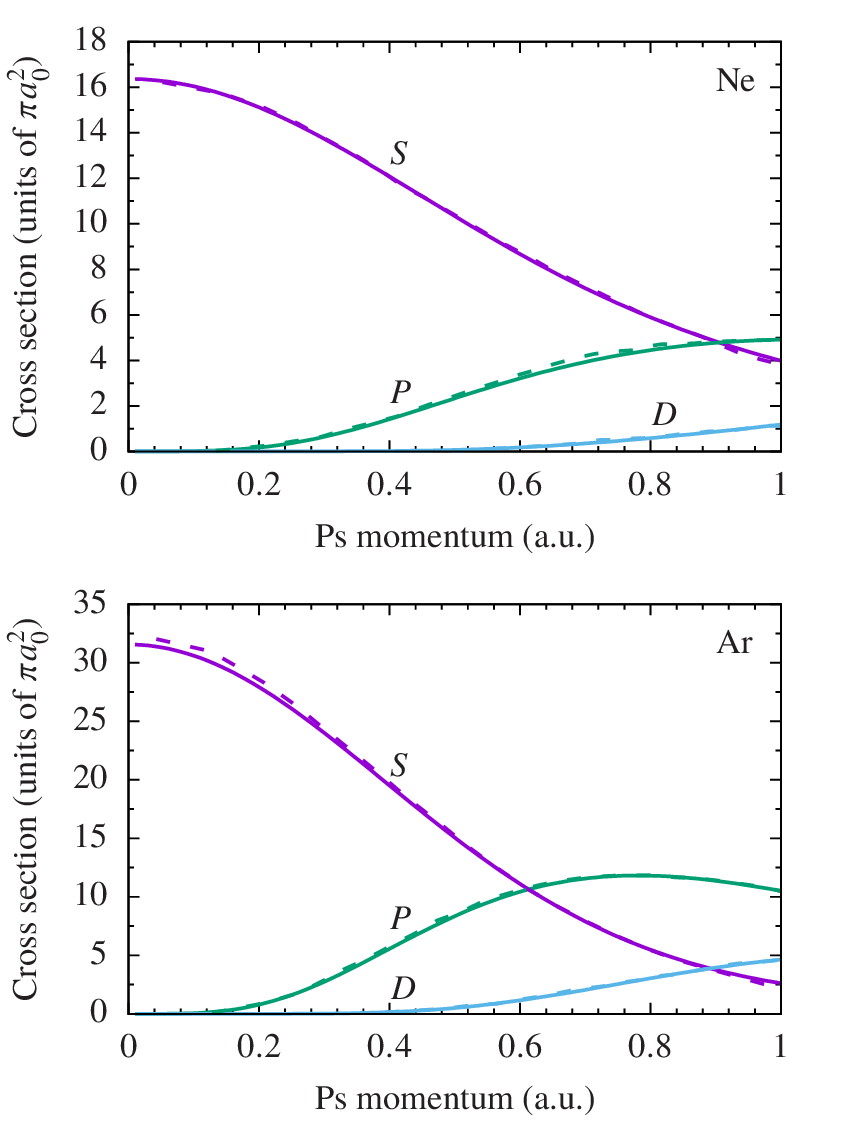}
\caption{\label{fig:NeAr_partial_cs}Partial cross sections for $S$-, $P$-, and $D$-wave scattering of Ps on Ne and Ar in the frozen-target approximation. Solid curves, present calculations; dashed curves, $R$-matrix calculations of Blackwood \textit{et al.} \cite{Blackwood02}.}
\end{figure}

For a more detailed comparison, Fig.~\ref{fig:NeAr_partial_cs} shows our partial cross sections $\sigma_L$ ($L=0$, 1, 2) for Ne and Ar in the frozen-target approximation and the corresponding results of the $R$-matrix calculations by Blackwood \textit{et al.} \cite{Blackwood02}.
With the exception of $S$-wave scattering on Ar at low energies, the two sets of calculations are practically indistinguishable. For $S$-wave scattering on Ar, there is a few-percent discrepancy at low $K$, but the overall agreement is still very good. This is evidence of the accuracy of our method of extracting scattering phase shifts from the Ps energy levels in the cavity.

In our calculations it was not possible to calculate the $F$-wave scattering phase shifts at the same level of accuracy as those with $L\leq 2$
[due to the increase in the size of the Hamiltonian matrix, cf. Eq.~(\ref{eq:dimH})]. The calculations of Blackwood \textit{et al.} \cite{Blackwood02,Blackwood03} did include this and higher partial waves. The agreement that we observe between their results and ours for He, Ne, Ar, and Kr up to $K=1$~a.u. indicates that the contribution from the $L\geq3$ partial waves is negligible in the considered energy range for these atoms. For Xe, the agreement is poorer at high energies, which may be an indication that the contribution of the $F$-wave is more important here.

For Ar and Xe, Fig.~\ref{fig:NeArKrXe_cs} also shows the experimental data of Brawley \textit{et al.} \cite{Brawley15}. These have a qualitatively different behaviour, with the cross section becoming smaller at lower momenta. Note though that the higher-energy data ($K>0.8$~a.u.) are closer to the calculated values.

\begin{figure*}
\includegraphics{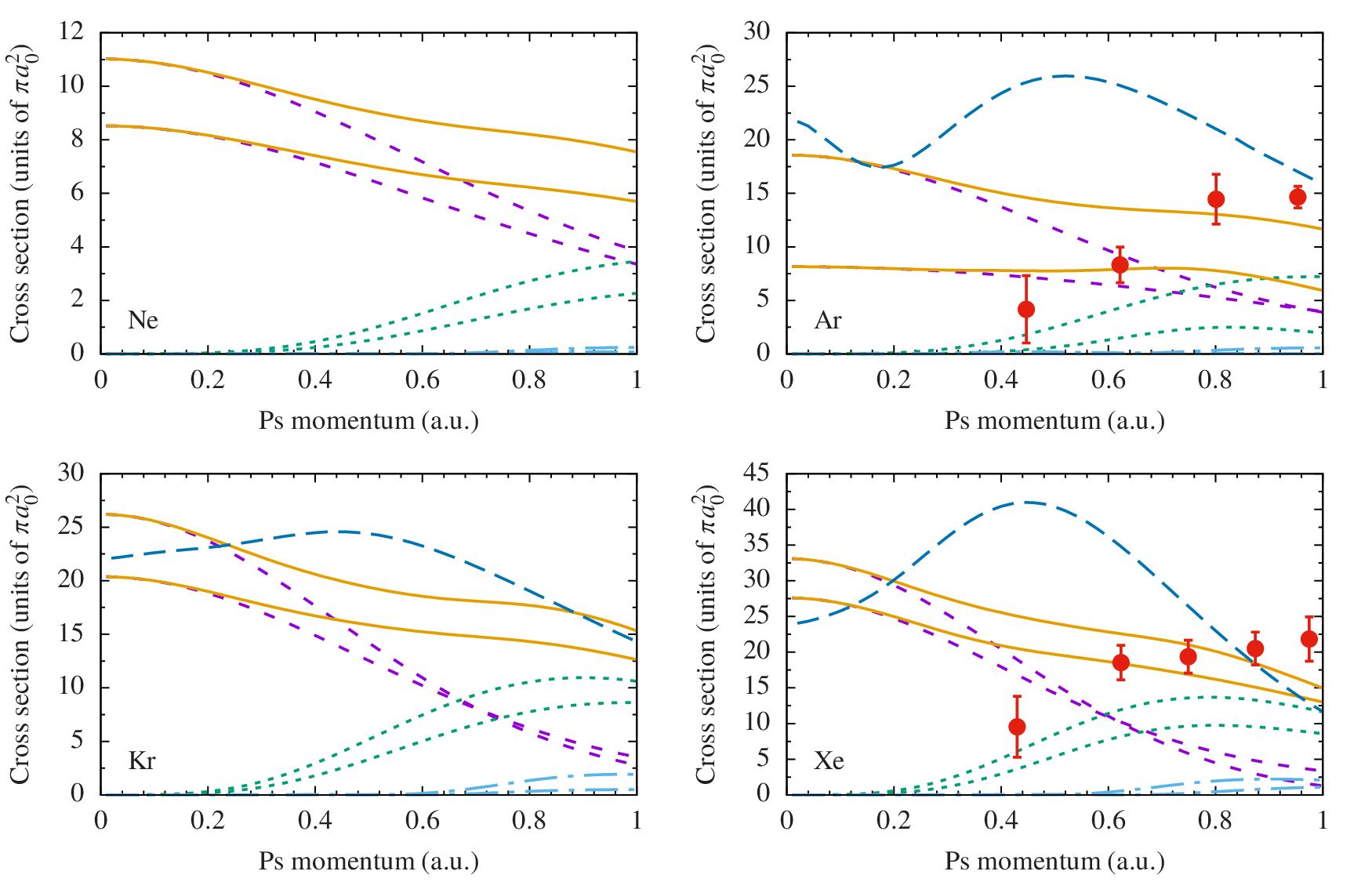}
\caption{\label{fig:NeArKrXe_cs_vdw}Partial and total elastic cross sections for Ps scattering on Ne, Ar, Kr, and Xe with inclusion of the van der Waals interaction. Short-dashed purple lines, $\sigma_0$; dotted green lines, $\sigma_1$; dash-dotted blue lines, $\sigma_2$; solid orange lines, $\sigma=\sigma_0+\sigma_1+\sigma_2$. For each cross section, curves that are higher (lower) at small $K$ correspond to the higher (lower) value of $R_0$. Long-dashed navy lines, pseudopotential calculations of $\sigma$ \cite{Fabrikant14,Gribakin16} (all for $R_0=3.0$~a.u.); filled red circles, experimental data \cite{Brawley15}.}
\end{figure*}

Figure~\ref{fig:NeArKrXe_cs_vdw} shows the total and partial elastic cross sections for Ps scattering on Ne, Ar, Kr, and Xe with inclusion of the van der Waals interaction, using two cutoff radii for each atom.
After the inclusion of the van der Waals interaction, the present cross sections retain the basic shape of their frozen-target counterparts in Fig.~\ref{fig:NeArKrXe_cs}, i.e., featureless and generally slowly decreasing. The relative decrease of each zero-energy cross section from its frozen-target value lies in the range 32--48\% for all target atoms and cutoff radii, except for Ar with $R_0=2.5$~a.u., for which the decrease is 74\%. This, and the earlier examination of the scattering lengths for Ar in Table~\ref{tab:NeArKrXe_data}, indicates that this cutoff radius is too small for Ar and the cross section for $R_0=3.0$~a.u. is more physical. We note that the additional attraction due to the van der Waals interaction, which counters the repulsive short-range static atomic potential, suppresses the contribution of higher partial waves in the energy range considered; indeed, the $L>1$ partial waves are essentially negligible for Ne and Ar.

There is a significant difference between our cross sections and the pseudopotential calculations that used the same van der Waals potential \cite{Fabrikant14,Gribakin16}. The present calculations do not support the appearance of broad maxima in Ar, Kr and Xe, or the low-$K$ minimum for Ar. We believe that these differences are due to the approximate treatment of the Ps interaction with the static atomic field by the pseudopotential method, which particularly affects the $P$ wave. We also note that inclusion of the van der Waals interaction does not resolve the discrepancy between the calculated cross sections and the measurements by Brawley \textit{et al.} \cite{Brawley15} for Ar and Xe.

\begin{figure*}
\includegraphics{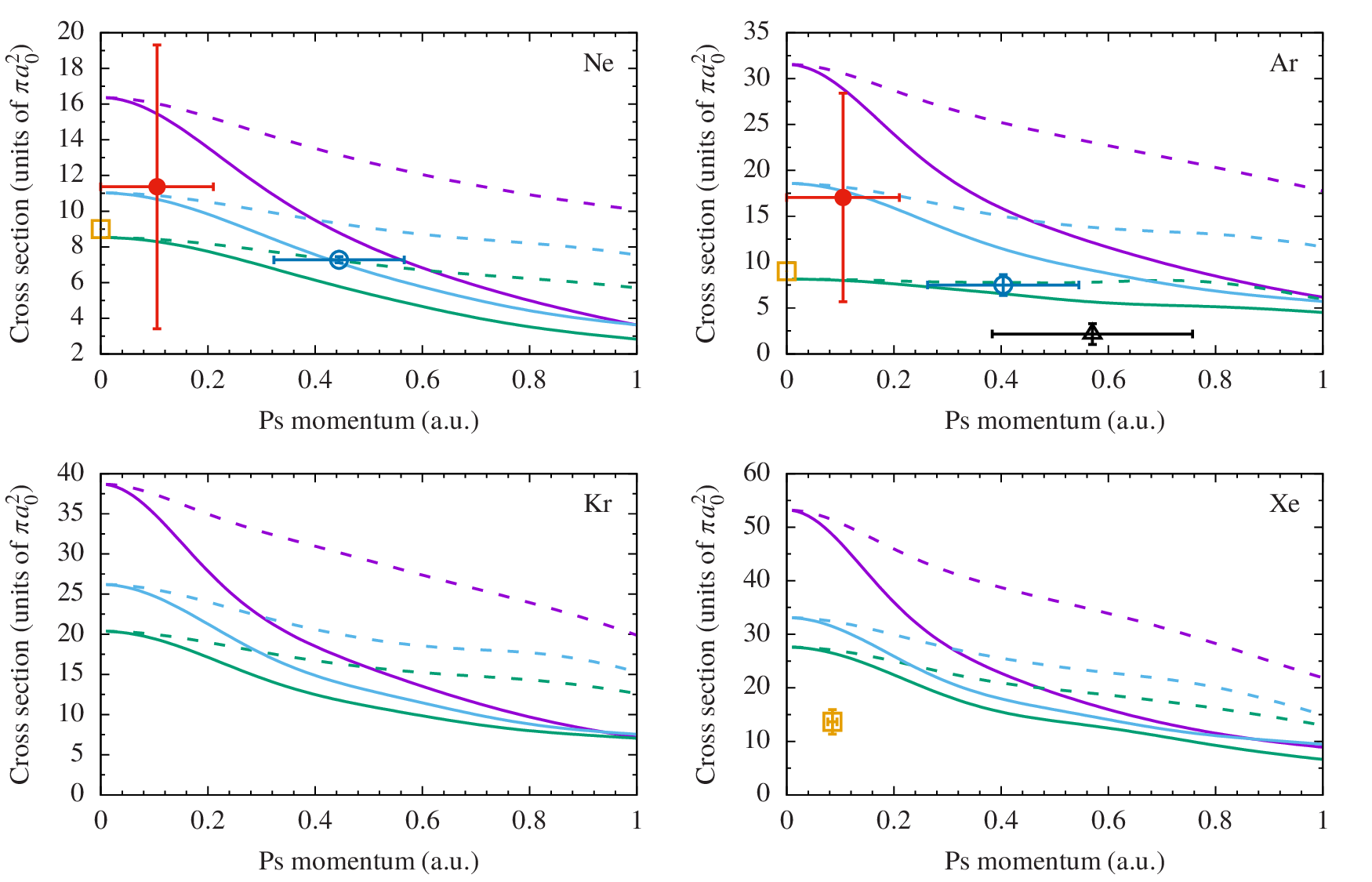}
\caption{\label{fig:PsNeArKrXe_MTCS}Calculated momentum-transfer cross sections (solid lines) and elastic cross sections (dashed lines) for Ps scattering on Ne, Ar, Kr, and Xe. In the order of decreasing magnitude: purple lines, frozen-target approximation; cyan lines, with the van der Waals interaction ($R_0=3.0$~a.u. for Ne and Ar, $R_0=3.5$~a.u. for Kr and Xe); green lines, with the van der Waals interaction ($R_0=2.5$~a.u. for Ne and Ar, $R_0=3.0$~a.u. for Kr and Xe). Measured momentum-transfer cross sections for Ne: open orange square, Coleman \textit{et al.} \cite{Coleman94}; open navy circle, Skalsey \textit{et al.} \cite{Skalsey03}; filled red circle, Nagashima \textit{et al.} \cite{Nagashima01}; for Ar: open orange square, Coleman \textit{et al.} \cite{Coleman94}; open navy circle, Skalsey \textit{et al.} \cite{Skalsey03}; filled red circle, Nagashima \textit{et al.} \cite{Nagashima95}; open black triangle, Sano \textit{et al.} \cite{Sano15}; for Xe: open orange square, Shibuya \textit{et al.} \cite{Shibuya13a}.}
\end{figure*}

Figure~\ref{fig:PsNeArKrXe_MTCS} shows the calculated momentum-transfer cross sections along with experimental data; also shown are our elastic cross sections.
The level of agreement between our calculations and the experimental data is mixed. For Ne, the data of Skalsey \textit{et al.} \cite{Skalsey03} and Nagashima \textit{et al.} \cite{Nagashima01} are in closest agreement with our van der Waals calculation with $R_0=3.0$~a.u., though both have large error bars. The zero-energy cross section of Coleman \textit{et al.} \cite{Coleman94} is in better agreement with our curve for $R_0=2.5$~a.u. For Ar, the measurements of Coleman \textit{et al.} \cite{Coleman94} and Skalsey \textit{et al.} \cite{Skalsey03} agree with our curve for $R_0=2.5$~a.u., and the value from Nagashima \textit{et al.} \cite{Nagashima95} is closer our curve for $R_0=3.5$~a.u. The recent measurement by Sano \textit{et al.} \cite{Sano15} is much lower than any of our calculations, and seems to contradict the earlier experimental data.

There are no measurements of the momentum-transfer cross section for Kr. For Xe, the only measurement is by Shibuya \textit{et al.} \cite{Shibuya13a}; their value
is about 50\% lower than the smallest of our theoretical predictions. Such a large difference could be at least partly due to the way in which the authors of Ref. \cite{Shibuya13a} deduced this datum. They compared the measured time evolution of the Ps center-of-mass energy $E(t)$ with the classical model described in Refs. \cite{Sauder68,Nagashima95a}, viz.,
\begin{equation}
E(t) = E_\text{th} \coth^2(\alpha+\beta t),
\end{equation}
where $t$ is the time, $E_\text{th}=E(\infty)=\frac32 k_B T$ is the thermal energy at temperature $T$ (with $k_B$ the Boltzmann constant), $\alpha=\coth^{-1}\sqrt{E(0)/E_\text{th}}$, $\beta=p_\text{th}\sigma_m n/M$, $p_\text{th}$ is the thermalized Ps momentum, $n$ is the gas number density, and $M$ is the mass of the atom (Xe). The values of $\alpha$ and $\beta$ were obtained from a least-squares fit of the measured $E(t)$ as $\alpha=0.094$ and $\beta=0.011$~ns$^{-1}$. However, their fit was based only on the restricted set of data ($40\leq E(t) \leq 60$~meV). Making use of all of the data points for $40\leq E(t) \leq 150$~meV, we find $\alpha=0.531$ and $\beta=0.0185$~ns$^{-1}$, giving a momentum-transfer cross section of $\sigma_m\approx 28\pi a_0^2$ at a Ps momentum of $K\sim 0.1$~a.u. This is a very significant increase from the value of $14\pi a_0^2$ quoted in Ref. \cite{Shibuya13}, and it would be in 10\% agreement with both of our van der Waals calculations.


\section{\label{sec:conc}Conclusions}

A new method has been developed to study Ps-atom interactions and has been applied to Ps scattering from the noble gases (He, Ne, Ar, Kr, and Xe). The main idea of the method is to calculate the states of Ps in a hard-wall cavity with the atom at the center. To implement this, a $B$-spline basis was used to construct single-particle electron and positron states in the static field of the target atom, with the entire system enclosed by an impenetrable spherical wall. The Ps wave function was obtained  as an expansion in the single-particle basis states. Diagonalization of the Hamiltonian matrix allowed the Ps energy levels to be found. From these the scattering phase shifts were determined, and scattering lengths and cross sections were computed for each target atom.

Comparisons were made with existing static-exchange, $R$-matrix, $T$-matrix, variational, and pseudopotential calculations. Scattering lengths were found to be in agreement with the fixed-core stochastic-variational results of Mitroy and coworkers \cite{Mitroy01,Mitroy03}, and partial elastic cross sections matched the results of Blackwood \textit{et al.} \cite{Blackwood02,Blackwood03} at the same level of approximation. Both comparisons attest to the accuracy of our method and its suitability for studying Ps-atom interactions.

The calculations were repeated with the long-range van der Waals interaction included by means of a model potential with an adjustable short-range cutoff parameter $R_0$. This allowed us to investigate the effect of dispersive forces on Ps-atom scattering. In most cases, the results were not very sensitive to the choice of the cutoff radius for physically motivated values of $R_0$
(close to the sum of the atomic and Ps radii).
For He, good accord was found with a previous calculation \cite{Walters04} that accounted for virtual excitation of both the Ps and He atoms. Agreement for the other target atoms with stochastic-variational \cite{Mitroy01,Mitroy03} (for scattering lengths) and pseudopotential \cite{Fabrikant14,Gribakin16} (for cross sections) calculations was varied. This was attributed to different ways of treating the dispersive interaction (for the former) or the static Ps-atom interaction (for the latter). Nevertheless, we were able to establish recommended values of the Ps-atom scattering lengths and zero-energy cross sections, with error bars reflecting the above discrepancies.


Comparisons with experiment were possible at the level of momentum-transfer cross sections inferred from Ps thermalization (typically at $K<0.5$~a.u.), and elastic scattering cross sections for Ar and Xe for $0.4< K<1$~a.u. For He, Ne, and Ar, some agreement was observed with a range of momentum-transfer data at low energies, partly assisted by large experimental error bars and some uncertainties in the calculations, related to the choice of $R_0$. Our calculations of the elastic cross sections for Ar and Xe disagree with the experimental observation of the cross section becoming very small at low Ps energies \cite{Brawley15}. The discrepancy is beyond the theoretical uncertainty due to the choice of $R_0$, and effectively rules out the possibility of a Ramsauer-Townsend-type minimum (a conclusion also supported by the study of the phase-shift behavior).

The general trends we have observed are as follows. At the level of the static (frozen-target) approximation, the Ps-atom interaction is repulsive. It is more repulsive for heavier (i.e., larger) atoms, producing the cross sections that increase along the noble-gas-atom sequence. Thus, the cross section for Ne is only slightly larger that that for He, while the the cross sections for Ar and Kr are greater by a factor of 2, and for Xe, by a factor of 3. Adding the attractive van der Waals interaction moderately reduces the Ps-atom repulsion. As a result, the cross sections become smaller by 30--50\%, with the effect being slightly larger for the heaver atoms, which have larger $C_6$ constants. It is interesting to contrast this with the role of correlation effects (i.e., polarization and virtual Ps formation) in low-energy positron-atom scattering, where the cross sections change dramatically, both qualitatively and quantitatively, in some cases by orders of magnitude \cite{Dzuba96,Green14}.

Looking ahead, the present method opens a new way for including correlation effects in Ps-atom interaction in a systematic \textit{ab initio} manner. This can be done by using many-body theory, which allows one to study properties of small 
and large atoms (e.g., He and Xe) with a similar level of accuracy (see, e.g., Refs. \cite{Gribakin04,Green13,Green14,Green15}, where this has been implemented for positron-atom and positron-ion collisions and annihilation). Such treatment should overcome the deficiencies of the model-van-der-Waals-potential approach and should enable one to calculate not only the scattering cross sections but also the Ps-atom pickoff annihilation rates.


\begin{acknowledgments}
We are grateful to G. Laricchia for communicating experimental data ahead of publication and to I. I. Fabrikant for useful discussions. The work of A.R.S. has been supported by the Department for the Economy, Northern Ireland.
\end{acknowledgments}

\appendix

\section{\label{app_vdw}Calculation of the matrix elements of the model van der Waals potential}

To calculate the matrix element $\langle \nu' \mu' \vert V_W \vert \mu \nu \rangle$ of the model van der Waals potential, Eq.~(\ref{eq:m_el_VW}), we first expand the potential (\ref{eqn:vdw_pot}) in Legendre polynomials $P_l$:
\begin{equation}\label{eq:a35}
V_W (R)
=
\sum_{l=0}^\infty \frac{[l]}{4\pi} V_W^{(l)}(r_e,r_p) P_l(\cos\omega),
\end{equation}
where $[l]\equiv2l+1$, $\omega$ is the angle between $\vec{r}_e$ and $\vec{r}_p$, and the expansion coefficients $V_W^{(l)}$ are to be determined. Multiplying both sides of Eq.~(\ref{eq:a35}) by $P_{l'}(\cos\omega)\sin\omega$, integrating over $\omega$ from $0$ to $\pi$, and changing variables to $x\equiv\cos\omega$, we obtain
\begin{align}
V_W^{(l)}(r_e,r_p) &= 2\pi \int_{-1}^1 V_W (R) P_l(x) \, dx \nonumber\\
&= -2^7 \pi C_6 \int_{-1}^1 \frac{ P_l(x)}{(r_e^2+r_p^2+2r_er_px)^3} \nonumber\\
&\quad{}\times\left\{ 1-\exp\left[ -\frac{(r_e^2+r_p^2+2r_er_px)^4}{(2R_0)^8}\right]\right\} \, dx,
\end{align}
where $R=\lvert \mathbf{r}_e+\mathbf{r}_p\rvert/2=(r_e^2+r_p^2+2 r_e r_p x)^{1/2}/2$.
These integrals are evaluated numerically using Gauss-Legendre quadrature with 40 abscissae. 
Factorizing the electron and positron wave functions as
\begin{equation}
\varphi_\mu(\mathbf{r}_e) = \frac{1}{r_e} P_\mu(r_e) Y_{l_\mu m_\mu}(\Omega_e)
\end{equation}
etc., 
carrying out the integration over the angular variables in the matrix element analytically, and assuming that the electron and positron are coupled to a total angular momentum $J$, we obtain an expression similar in form to Eq.~(A3) of Ref. \cite{Brown17}, viz.,
\begin{align}
\langle \nu' \mu' \vert V_W \vert \mu \nu \rangle = \sum_{l} (-1)^{J+l}
\begin{Bmatrix}
J & l_{\mu'} & l_{\nu'} \\ l & l_\nu & l_\mu
\end{Bmatrix}
\langle \nu'\mu' \Vert V_W^{(l)} \Vert \mu \nu \rangle,
\end{align}
where
\begin{align}\label{eq:a5}
\langle \nu'\mu' \Vert V_W^{(l)} \Vert \mu \nu \rangle &= 
\frac{[l]}{4\pi}
\sqrt{[l_{\nu'}][l_{\mu'}][l_\mu][l_\nu]}
\begin{pmatrix}
l_{\mu'} & l & l_\mu \\ 0 & 0 & 0
\end{pmatrix}
\begin{pmatrix}
l_{\nu'} & l & l_\nu \\ 0 & 0 & 0
\end{pmatrix}
\nonumber\\
&\quad{}\times
\int_0^{R_c}\!\!\int_0^{R_c}  P_{\nu'}(r_p) P_{\mu'}(r_e) V_W^{(l)}(r_e,r_p) \nonumber\\
&\quad{}\times P_{\mu}(r_e) P_{\nu}(r_p) \, dr_e \, dr_p.
\end{align}
The double radial integral in Eq.~(\ref{eq:a5}) is evaluated numerically.

\section{\label{sec:coll_rad}Collisional radius of positronium}

In Ref. \cite{Brown17} the dependence of the collisional radius $\rho(K)$ of Ps($1s$) on the Ps center-of-mass momentum $K$ was given as a linear fit, viz.,
\begin{equation}
\rho(K)=1.65-0.51K.
\end{equation}
This simple fit was determined by calculating Ps energy levels in an otherwise empty cavity with $R_c=10$ and 12~a.u. It gives a good overall description of the momentum dependence of $\rho (K)$. However, it is not sufficiently accurate for extracting the Ps-atom phase shifts from the bound-state energies, as described by Eq.~(\ref{eq:phashft_form}). Hence, in the present work we have used a new set of more flexible, nonlinear fits for $\rho(K)$, considering each combination of $R_c$ and $J^\Pi$ separately. 
%
%

For Ps confined in a spherical cavity, it is possible that $\rho$ (defined as explained in Ref.~\cite{Brown17}) may have some dependence on the cavity radius $R_c$ (for a given momentum $K$). Thus, if $R_c$ is small then the cavity wall has a large curvature, meaning that the Ps interacts with it more strongly and is held further from it, resulting in a large value of $\rho$. Conversely, if $R_c$ is large then the cavity wall is flatter, and the Ps can approach it more closely, resulting in a smaller $\rho$. There may also be some inaccuracies in the representation of Ps close to the wall (where the single-center expansion of the Ps wave function is slower to converge), which can be effectively absorbed into the value of $\rho (K)$.
For this reason, we have generated a separate fit for each combination of $R_c$ and $J^\Pi$.

The precise form of each fit depends on the number of Ps energy eigenstates (and corresponding values of $\rho $) available for given $R_c$ and $J^\Pi$ (see Ref.~\cite{Brown17}); this number varies between two and four. If only two data points are available, we resort to a linear fit
\begin{equation}\label{eq:2par}
\rho (K)=b_0+b_1K;
\end{equation}
if three are available, we use the Pad\'e approximant
\begin{equation}\label{eq:3par}
\rho(K) = \frac{b_0 + b_1 K^2}{1 + b_2 K};
\end{equation}
and if four are available, we use
\begin{equation}\label{eq:4par}
\rho(K) = \frac{b_0 + b_1 K^2 + b_2 K^4}{1 + b_3 K^3},
\end{equation}
where the $b_j$ are the fitting parameters listed in Table~\ref{tab:rad_fits_b}.


\begin{table}
\caption{\label{tab:rad_fits_b}Parameters of fits for $\rho(K)$ for different cavity radii and symmetries. The number of parameters in each line indicates which of the three fits [Eq.~(\ref{eq:2par}), (\ref{eq:3par}), or (\ref{eq:4par})] was used.}
\begin{ruledtabular}
\begin{tabular}{cccccc}
$J^\Pi$ & $R_c$ (a.u.) & $b_0$ & $b_1$ & $b_2$ & $b_3$ \\
\hline
$0^+$ & 10 & 1.67811 & $-0.130277$ & 0.360395 &  \\
      & 12 & 1.44687 & 0.112604  & 0.475122 & 0.823573 \\
      & 14 & 1.34200 & $-0.502768$ & $-0.175790$ &  \\
      & 16 & 1.25292 & 2.51017 & 1.37486 & 3.69569 \\
 \hline          
$1^-$ & 10 & 1.7586 & $-0.150502$ & 0.423148 &   \\
      & 12 & 1.59782 & $-0.249293$ & 0.228386 &  \\
      & 14 & 1.50500 & $-0.326394$ & 0.102026 &  \\
      & 16 & 1.31856 & $-0.580538$ & $-0.247459$ &  \\
 \hline         
$2^+$ & 10 & 1.79929 & $-0.649781$ &  &  \\
      & 12 & 1.88632 & $-0.172525$ & 0.510590 &   \\
      & 14 & 2.31747 & $-0.00086334$ & 1.02952 &  \\
      & 16 & 2.23371 & 0.134734 & 1.09300 &                 
\end{tabular}
\end{ruledtabular}
\end{table}

\section{\label{sec:tables}Effective-range-theory fits for the scattering phase shifts}

The effective-range-type fits used for Ps-atom scattering phase shifts in the frozen-target approximation are
\begin{subequations}\label{eq:d}
\begin{align}
\delta_0(K) &= a_0 K + a_1 K^3 + a_2 K^5, \label{eq:d0} \\
\delta_1(K) &= \frac{a_0 K^3}{1 + a_1K^2 + a_2K^4},\label{eq:d1}\\
\delta_2(K) &= \frac{a_0 K^5}{1 + a_1K^2 + a_2K^4 + a_3K^6}.\label{eq:d2}
\end{align}
\end{subequations}
When the van der Waals interaction is included, we use
\begin{subequations}
\begin{align}
\delta_0 &= a_0K + a_1 K^3, \label{eq:d0_vdW}\\
\delta_1 &= \frac{a_0K^3}{1+a_1K^2+a_2K^4}, \label{eq:d1_vdW}\\
\delta_2 &= \frac{a_0K^4 + a_1K^5}{1+a_2K^6}.\label{eq:d2_vdW}
\end{align}
\end{subequations}
Tables~\ref{tab:c1}--\ref{tab:c3} list the values of the fitting parameters $a_j$ for each target atom.

\begin{table}
\caption{\label{tab:c1}Parameters of the fits for the $S$-, $P$-, and $D$-wave phase shifts $\delta_L$ for Ps scattering on noble-gas atoms in the frozen-target approximation.}
\begin{ruledtabular}
\begin{tabular}{ccD{.}{.}{-1.3}D{.}{.}{-1.3}D{.}{.}{-1.4}D{.}{.}{-2.5}D{.}{.}{-2.3}}
&& \multicolumn{5}{c}{Atom} \\
\cline{3-7}
$L$ & $a_j$ & \multicolumn{1}{c}{He} & \multicolumn{1}{c}{Ne} & \multicolumn{1}{c}{Ar} & \multicolumn{1}{c}{Kr} & \multicolumn{1}{c}{Xe}  \\
\hline
$0$ & $a_0$ & -1.86 & -2.02 & -2.81 & -3.11 & -3.65 \\
       & $a_1$ & -2.13 & 0.656 & 0.634 & 0.663 & 1.31 \\
       & $a_2$ & -0.781 & -0.141 & -0.0249 & -0.00564 & -0.708 \\ 
\hline
$1$ & $a_0$ & -2.13 & -3.41 & -7.62 & -11.1 & -14.6 \\
       & $a_1$ & 2.57 & 3.61 & 4.71 & 6.30 & 6.79 \\
       & $a_2$ & 1.94 & 0.310 & 0.588 & 0.503 & 0.698 \\
\hline
$2$ & $a_0$ & -0.781 & -1.43 & -6.45 & -14.5 & -15.8 \\
       & $a_1$ & 2.57 & 1.17 & 4.68 & 11.4 & 5.05 \\
       & $a_2$ & 2.16 & 4.80 & 5.38 & 2.55 & 11.7 \\
       & $a_3$ & 2.75 & -1.16 & 1.77 & 8.92 & 2.21 
\end{tabular}
\end{ruledtabular}
\end{table}

\begin{table}
\caption{\label{tab:c2}Parameters of the fits for the $S$-, $P$-, and $D$-wave phase shifts $\delta_L$ for Ps scattering on noble-gas atoms with inclusion of the van der Waals interaction using the lower cutoff radius ($R_0=2.5$~a.u. for He, Ne, and Ar, and $R_0=3.0$~a.u. for Kr and Xe).}
\begin{ruledtabular}
\begin{tabular}{ccD{.}{.}{-1.4}D{.}{.}{-1.3}D{.}{.}{-2.4}D{.}{.}{-1.3}D{.}{.}{-1.3}}
&& \multicolumn{5}{c}{Atom} \\
\cline{3-7}
$L$ & $a_j$ & \multicolumn{1}{c}{He} & \multicolumn{1}{c}{Ne} & \multicolumn{1}{c}{Ar} & \multicolumn{1}{c}{Kr} & \multicolumn{1}{c}{Xe}  \\
\hline
$0$ & $a_0$ & -1.52 & -1.46 & -1.43 & -2.26 & -2.63 \\
       & $a_1$ & 0.299 & 0.303 & -0.0414 & 0.332 & 0.644 \\
\hline
$1$ & $a_0$ & -0.662 & -1.01 & -0.739 & -3.22 & -3.89 \\
       & $a_1$ & -0.0664 & 0.822 & -1.30 & 2.01 & 1.30 \\
       & $a_2$ & 2.36 & 0.432 & 2.08 & 0.175 & 1.57 \\
\hline
$2$ & $a_0$ & 0.778 & 1.12 & -1.03 & 2.23 & 2.27 \\
       & $a_1$ & -0.850 & -1.81 & 11.9 & -3.89 & -3.67 \\
       & $a_2$ & 76.6 & 9.92 & 351 & 9.53 & 4.97 
\end{tabular}
\end{ruledtabular}
\end{table}

\begin{table}
\caption{\label{tab:c3}Parameters of the fits for the $S$-, $P$-, and $D$-wave phase shifts $\delta_L$ for Ps scattering on noble-gas atoms with inclusion of the van der Waals interaction using the higher cutoff radius ($R_0=3.0$~a.u. for He, Ne, and Ar, and $R_0=3.5$~a.u. for Kr and Xe).}
\begin{ruledtabular}
\begin{tabular}{ccD{.}{.}{-1.3}D{.}{.}{-1.3}D{.}{.}{-1.3}D{.}{.}{-1.4}D{.}{.}{-1.3}}
&& \multicolumn{5}{c}{Atom} \\
\cline{3-7}
$L$ & $a_j$ & \multicolumn{1}{c}{He} & \multicolumn{1}{c}{Ne} & \multicolumn{1}{c}{Ar} & \multicolumn{1}{c}{Kr} & \multicolumn{1}{c}{Xe}  \\
\hline
$0$ & $a_0$ & -1.61 & -1.66 & -2.16 & -2.56 & -2.88 \\
       & $a_1$ & 0.265 & 0.289 & 0.418 & 0.407 & 0.333 \\
\hline
$1$ & $a_0$ & -1.27 & -1.47 & -2.46 & -4.41 & -6.07 \\
       & $a_1$ & 1.78 & 1.19 & 1.20 & 2.59 & 2.89 \\
       & $a_2$ & 1.36 & 0.416 & 0.573 & 0.0192 & 0.464 \\
\hline
$2$ & $a_0$ & 0.303 & 0.805 & 1.61 & 1.50 & 1.95 \\
       & $a_1$ & -0.555 & -1.47 & -2.87 & -3.31 & -4.70 \\
       & $a_2$ & 6.81 & 5.01 & 6.47 & 4.73 & 7.38 
\end{tabular}
\end{ruledtabular}
\end{table}


\begin{thebibliography}{86}%
\makeatletter
\providecommand \@ifxundefined [1]{%
 \@ifx{#1\undefined}
}%
\providecommand \@ifnum [1]{%
 \ifnum #1\expandafter \@firstoftwo
 \else \expandafter \@secondoftwo
 \fi
}%
\providecommand \@ifx [1]{%
 \ifx #1\expandafter \@firstoftwo
 \else \expandafter \@secondoftwo
 \fi
}%
\providecommand \natexlab [1]{#1}%
\providecommand \enquote  [1]{``#1''}%
\providecommand \bibnamefont  [1]{#1}%
\providecommand \bibfnamefont [1]{#1}%
\providecommand \citenamefont [1]{#1}%
\providecommand \href@noop [0]{\@secondoftwo}%
\providecommand \href [0]{\begingroup \@sanitize@url \@href}%
\providecommand \@href[1]{\@@startlink{#1}\@@href}%
\providecommand \@@href[1]{\endgroup#1\@@endlink}%
\providecommand \@sanitize@url [0]{\catcode `\\12\catcode `\$12\catcode
  `\&12\catcode `\#12\catcode `\^12\catcode `\_12\catcode `\%12\relax}%
\providecommand \@@startlink[1]{}%
\providecommand \@@endlink[0]{}%
\providecommand \url  [0]{\begingroup\@sanitize@url \@url }%
\providecommand \@url [1]{\endgroup\@href {#1}{\urlprefix }}%
\providecommand \urlprefix  [0]{URL }%
\providecommand \Eprint [0]{\href }%
\providecommand \doibase [0]{http://dx.doi.org/}%
\providecommand \selectlanguage [0]{\@gobble}%
\providecommand \bibinfo  [0]{\@secondoftwo}%
\providecommand \bibfield  [0]{\@secondoftwo}%
\providecommand \translation [1]{[#1]}%
\providecommand \BibitemOpen [0]{}%
\providecommand \bibitemStop [0]{}%
\providecommand \bibitemNoStop [0]{.\EOS\space}%
\providecommand \EOS [0]{\spacefactor3000\relax}%
\providecommand \BibitemShut  [1]{\csname bibitem#1\endcsname}%
\let\auto@bib@innerbib\@empty
\bibitem [{\citenamefont {Laricchia}\ and\ \citenamefont
  {Walters}(2012)}]{Laricchia12}%
  \BibitemOpen
  \bibfield  {author} {\bibinfo {author} {\bibfnamefont {G.}~\bibnamefont
  {Laricchia}}\ and\ \bibinfo {author} {\bibfnamefont {H.~R.~J.}\ \bibnamefont
  {Walters}},\ }\href {\doibase 10.1393/ncr/i2012-10077-6} {\bibfield
  {journal} {\bibinfo  {journal} {Riv. Nuovo. Cimento}\ }\textbf {\bibinfo
  {volume} {35}},\ \bibinfo {pages} {305} (\bibinfo {year} {2012})}\BibitemShut
  {NoStop}%
\bibitem [{\citenamefont {Kellerbauer}\ \emph {et~al.}(2008)\citenamefont
  {Kellerbauer}, \citenamefont {Amoretti}, \citenamefont {Belov}, \citenamefont
  {Bonomi}, \citenamefont {Boscolo}, \citenamefont {Brusa}, \citenamefont
  {B\"uchner}, \citenamefont {Byakov}, \citenamefont {Cabaret}, \citenamefont
  {Canali}, \citenamefont {Carraro}, \citenamefont {Castelli}, \citenamefont
  {Cialdi}, \citenamefont {de~Combarieu}, \citenamefont {Comparat},
  \citenamefont {Consolati}, \citenamefont {Djourelov}, \citenamefont {Doser},
  \citenamefont {Drobychev}, \citenamefont {Dupasquier}, \citenamefont
  {Ferrari}, \citenamefont {Forget}, \citenamefont {Formaro}, \citenamefont
  {Gervasini}, \citenamefont {Giammarchi}, \citenamefont {Gninenko},
  \citenamefont {Gribakin}, \citenamefont {Hogan}, \citenamefont {Jacquey},
  \citenamefont {Lagomarsino}, \citenamefont {Manuzio}, \citenamefont
  {Mariazzi}, \citenamefont {Matveev}, \citenamefont {Meier}, \citenamefont
  {Merkt}, \citenamefont {Nedelec}, \citenamefont {Oberthaler}, \citenamefont
  {Pari}, \citenamefont {Prevedelli}, \citenamefont {Quasso}, \citenamefont
  {Rotondi}, \citenamefont {Sillou}, \citenamefont {Stepanov}, \citenamefont
  {Stroke}, \citenamefont {Testera}, \citenamefont {Tino}, \citenamefont
  {Tr\'enec}, \citenamefont {Vairo}, \citenamefont {Vigu\'e}, \citenamefont
  {Walters}, \citenamefont {Warring}, \citenamefont {Zavatarelli},\ and\
  \citenamefont {Zvezhinskij}}]{Kellerbauer08}%
  \BibitemOpen
  \bibfield  {author} {\bibinfo {author} {\bibfnamefont {A.}~\bibnamefont
  {Kellerbauer}}, \bibinfo {author} {\bibfnamefont {M.}~\bibnamefont
  {Amoretti}}, \bibinfo {author} {\bibfnamefont {A.}~\bibnamefont {Belov}},
  \bibinfo {author} {\bibfnamefont {G.}~\bibnamefont {Bonomi}}, \bibinfo
  {author} {\bibfnamefont {I.}~\bibnamefont {Boscolo}}, \bibinfo {author}
  {\bibfnamefont {R.}~\bibnamefont {Brusa}}, \bibinfo {author} {\bibfnamefont
  {M.}~\bibnamefont {B\"uchner}}, \bibinfo {author} {\bibfnamefont
  {V.}~\bibnamefont {Byakov}}, \bibinfo {author} {\bibfnamefont
  {L.}~\bibnamefont {Cabaret}}, \bibinfo {author} {\bibfnamefont
  {C.}~\bibnamefont {Canali}}, \bibinfo {author} {\bibfnamefont
  {C.}~\bibnamefont {Carraro}}, \bibinfo {author} {\bibfnamefont
  {F.}~\bibnamefont {Castelli}}, \bibinfo {author} {\bibfnamefont
  {S.}~\bibnamefont {Cialdi}}, \bibinfo {author} {\bibfnamefont
  {M.}~\bibnamefont {de~Combarieu}}, \bibinfo {author} {\bibfnamefont
  {D.}~\bibnamefont {Comparat}}, \bibinfo {author} {\bibfnamefont
  {G.}~\bibnamefont {Consolati}}, \bibinfo {author} {\bibfnamefont
  {N.}~\bibnamefont {Djourelov}}, \bibinfo {author} {\bibfnamefont
  {M.}~\bibnamefont {Doser}}, \bibinfo {author} {\bibfnamefont
  {G.}~\bibnamefont {Drobychev}}, \bibinfo {author} {\bibfnamefont
  {A.}~\bibnamefont {Dupasquier}}, \bibinfo {author} {\bibfnamefont
  {G.}~\bibnamefont {Ferrari}}, \bibinfo {author} {\bibfnamefont
  {P.}~\bibnamefont {Forget}}, \bibinfo {author} {\bibfnamefont
  {L.}~\bibnamefont {Formaro}}, \bibinfo {author} {\bibfnamefont
  {A.}~\bibnamefont {Gervasini}}, \bibinfo {author} {\bibfnamefont
  {M.}~\bibnamefont {Giammarchi}}, \bibinfo {author} {\bibfnamefont
  {S.}~\bibnamefont {Gninenko}}, \bibinfo {author} {\bibfnamefont
  {G.}~\bibnamefont {Gribakin}}, \bibinfo {author} {\bibfnamefont
  {S.}~\bibnamefont {Hogan}}, \bibinfo {author} {\bibfnamefont
  {M.}~\bibnamefont {Jacquey}}, \bibinfo {author} {\bibfnamefont
  {V.}~\bibnamefont {Lagomarsino}}, \bibinfo {author} {\bibfnamefont
  {G.}~\bibnamefont {Manuzio}}, \bibinfo {author} {\bibfnamefont
  {S.}~\bibnamefont {Mariazzi}}, \bibinfo {author} {\bibfnamefont
  {V.}~\bibnamefont {Matveev}}, \bibinfo {author} {\bibfnamefont
  {J.}~\bibnamefont {Meier}}, \bibinfo {author} {\bibfnamefont
  {F.}~\bibnamefont {Merkt}}, \bibinfo {author} {\bibfnamefont
  {P.}~\bibnamefont {Nedelec}}, \bibinfo {author} {\bibfnamefont
  {M.}~\bibnamefont {Oberthaler}}, \bibinfo {author} {\bibfnamefont
  {P.}~\bibnamefont {Pari}}, \bibinfo {author} {\bibfnamefont {M.}~\bibnamefont
  {Prevedelli}}, \bibinfo {author} {\bibfnamefont {F.}~\bibnamefont {Quasso}},
  \bibinfo {author} {\bibfnamefont {A.}~\bibnamefont {Rotondi}}, \bibinfo
  {author} {\bibfnamefont {D.}~\bibnamefont {Sillou}}, \bibinfo {author}
  {\bibfnamefont {S.}~\bibnamefont {Stepanov}}, \bibinfo {author}
  {\bibfnamefont {H.}~\bibnamefont {Stroke}}, \bibinfo {author} {\bibfnamefont
  {G.}~\bibnamefont {Testera}}, \bibinfo {author} {\bibfnamefont
  {G.}~\bibnamefont {Tino}}, \bibinfo {author} {\bibfnamefont {G.}~\bibnamefont
  {Tr\'enec}}, \bibinfo {author} {\bibfnamefont {A.}~\bibnamefont {Vairo}},
  \bibinfo {author} {\bibfnamefont {J.}~\bibnamefont {Vigu\'e}}, \bibinfo
  {author} {\bibfnamefont {H.}~\bibnamefont {Walters}}, \bibinfo {author}
  {\bibfnamefont {U.}~\bibnamefont {Warring}}, \bibinfo {author} {\bibfnamefont
  {S.}~\bibnamefont {Zavatarelli}}, \ and\ \bibinfo {author} {\bibfnamefont
  {D.}~\bibnamefont {Zvezhinskij}},\ }\href {\doibase
  http://dx.doi.org/10.1016/j.nimb.2007.12.010} {\bibfield  {journal} {\bibinfo
   {journal} {Nucl. Instrum. Methods B}\ }\textbf {\bibinfo {volume} {266}},\
  \bibinfo {pages} {351} (\bibinfo {year} {2008})}\BibitemShut {NoStop}%
\bibitem [{\citenamefont {Debu}(2012)}]{Debu2012}%
  \BibitemOpen
  \bibfield  {author} {\bibinfo {author} {\bibfnamefont {P.}~\bibnamefont
  {Debu}},\ }\href {\doibase 10.1007/s10751-011-0379-4} {\bibfield  {journal}
  {\bibinfo  {journal} {Hyperfine Interactions}\ }\textbf {\bibinfo {volume}
  {212}},\ \bibinfo {pages} {51} (\bibinfo {year} {2012})}\BibitemShut
  {NoStop}%
\bibitem [{\citenamefont {Consolati}\ \emph {et~al.}(2013)\citenamefont
  {Consolati}, \citenamefont {Ferragut}, \citenamefont {Galarneau},
  \citenamefont {{Di Renzo}},\ and\ \citenamefont {Quasso}}]{Consolati13}%
  \BibitemOpen
  \bibfield  {author} {\bibinfo {author} {\bibfnamefont {G.}~\bibnamefont
  {Consolati}}, \bibinfo {author} {\bibfnamefont {R.}~\bibnamefont {Ferragut}},
  \bibinfo {author} {\bibfnamefont {A.}~\bibnamefont {Galarneau}}, \bibinfo
  {author} {\bibfnamefont {F.}~\bibnamefont {{Di Renzo}}}, \ and\ \bibinfo
  {author} {\bibfnamefont {F.}~\bibnamefont {Quasso}},\ }\href {\doibase
  10.1039/c2cs35454c} {\bibfield  {journal} {\bibinfo  {journal} {Chem. Soc.
  Rev.}\ }\textbf {\bibinfo {volume} {42}},\ \bibinfo {pages} {3821} (\bibinfo
  {year} {2013})}\BibitemShut {NoStop}%
\bibitem [{\citenamefont {Gidley}\ \emph {et~al.}(2006)\citenamefont {Gidley},
  \citenamefont {Peng},\ and\ \citenamefont {Vallery}}]{Gidley06}%
  \BibitemOpen
  \bibfield  {author} {\bibinfo {author} {\bibfnamefont {D.~W.}\ \bibnamefont
  {Gidley}}, \bibinfo {author} {\bibfnamefont {H.}~\bibnamefont {Peng}}, \ and\
  \bibinfo {author} {\bibfnamefont {R.~S.}\ \bibnamefont {Vallery}},\ }\href
  {\doibase 10.1146/annurev.matsci.36.111904.135144} {\bibfield  {journal}
  {\bibinfo  {journal} {Riv. Nuovo. Cimento}\ }\textbf {\bibinfo {volume}
  {36}},\ \bibinfo {pages} {49} (\bibinfo {year} {2006})}\BibitemShut {NoStop}%
\bibitem [{\citenamefont {Shibuya}\ \emph
  {et~al.}(2013{\natexlab{a}})\citenamefont {Shibuya}, \citenamefont
  {Nakayama}, \citenamefont {Saito},\ and\ \citenamefont {Hyodo}}]{Shibuya13}%
  \BibitemOpen
  \bibfield  {author} {\bibinfo {author} {\bibfnamefont {K.}~\bibnamefont
  {Shibuya}}, \bibinfo {author} {\bibfnamefont {T.}~\bibnamefont {Nakayama}},
  \bibinfo {author} {\bibfnamefont {H.}~\bibnamefont {Saito}}, \ and\ \bibinfo
  {author} {\bibfnamefont {T.}~\bibnamefont {Hyodo}},\ }\href {\doibase
  10.1103/PhysRevA.88.012511} {\bibfield  {journal} {\bibinfo  {journal} {Phys.
  Rev. A}\ }\textbf {\bibinfo {volume} {88}},\ \bibinfo {pages} {012511}
  (\bibinfo {year} {2013}{\natexlab{a}})}\BibitemShut {NoStop}%
\bibitem [{\citenamefont {Shibuya}\ \emph
  {et~al.}(2013{\natexlab{b}})\citenamefont {Shibuya}, \citenamefont
  {Kawamura},\ and\ \citenamefont {Saito}}]{Shibuya13a}%
  \BibitemOpen
  \bibfield  {author} {\bibinfo {author} {\bibfnamefont {K.}~\bibnamefont
  {Shibuya}}, \bibinfo {author} {\bibfnamefont {Y.}~\bibnamefont {Kawamura}}, \
  and\ \bibinfo {author} {\bibfnamefont {H.}~\bibnamefont {Saito}},\ }\href
  {\doibase 10.1103/PhysRevA.88.042517} {\bibfield  {journal} {\bibinfo
  {journal} {Phys. Rev. A}\ }\textbf {\bibinfo {volume} {88}},\ \bibinfo
  {pages} {042517} (\bibinfo {year} {2013}{\natexlab{b}})}\BibitemShut
  {NoStop}%
\bibitem [{\citenamefont {Cassidy}\ and\ \citenamefont {{Mills,
  Jr.}}(2007{\natexlab{a}})}]{Cassidy07}%
  \BibitemOpen
  \bibfield  {author} {\bibinfo {author} {\bibfnamefont {D.~B.}\ \bibnamefont
  {Cassidy}}\ and\ \bibinfo {author} {\bibfnamefont {A.~P.}\ \bibnamefont
  {{Mills, Jr.}}},\ }\href {\doibase 10.1038/nature06094} {\bibfield  {journal}
  {\bibinfo  {journal} {Nature}\ }\textbf {\bibinfo {volume} {449}},\ \bibinfo
  {pages} {195} (\bibinfo {year} {2007}{\natexlab{a}})}\BibitemShut {NoStop}%
\bibitem [{\citenamefont {Cassidy}\ and\ \citenamefont {{Mills,
  Jr.}}(2008)}]{Cassidy08}%
  \BibitemOpen
  \bibfield  {author} {\bibinfo {author} {\bibfnamefont {D.~B.}\ \bibnamefont
  {Cassidy}}\ and\ \bibinfo {author} {\bibfnamefont {A.~P.}\ \bibnamefont
  {{Mills, Jr.}}},\ }\href {\doibase 10.1103/PhysRevLett.100.013401} {\bibfield
   {journal} {\bibinfo  {journal} {Phys. Rev. Lett.}\ }\textbf {\bibinfo
  {volume} {100}},\ \bibinfo {pages} {013401} (\bibinfo {year}
  {2008})}\BibitemShut {NoStop}%
\bibitem [{\citenamefont {Cassidy}\ and\ \citenamefont {{Mills,
  Jr.}}(2011)}]{Cassidy11}%
  \BibitemOpen
  \bibfield  {author} {\bibinfo {author} {\bibfnamefont {D.~B.}\ \bibnamefont
  {Cassidy}}\ and\ \bibinfo {author} {\bibfnamefont {A.~P.}\ \bibnamefont
  {{Mills, Jr.}}},\ }\href {\doibase 10.1103/PhysRevLett.107.213401} {\bibfield
   {journal} {\bibinfo  {journal} {Phys. Rev. Lett.}\ }\textbf {\bibinfo
  {volume} {107}},\ \bibinfo {pages} {213401} (\bibinfo {year}
  {2011})}\BibitemShut {NoStop}%
\bibitem [{\citenamefont {Cassidy}\ \emph {et~al.}(2012)\citenamefont
  {Cassidy}, \citenamefont {Hisakado}, \citenamefont {Tom},\ and\ \citenamefont
  {{Mills, Jr.}}}]{Cassidy12}%
  \BibitemOpen
  \bibfield  {author} {\bibinfo {author} {\bibfnamefont {D.~B.}\ \bibnamefont
  {Cassidy}}, \bibinfo {author} {\bibfnamefont {T.~H.}\ \bibnamefont
  {Hisakado}}, \bibinfo {author} {\bibfnamefont {H.~W.~K.}\ \bibnamefont
  {Tom}}, \ and\ \bibinfo {author} {\bibfnamefont {A.~P.}\ \bibnamefont
  {{Mills, Jr.}}},\ }\href {\doibase 10.1103/PhysRevLett.108.133402} {\bibfield
   {journal} {\bibinfo  {journal} {Phys. Rev. Lett.}\ }\textbf {\bibinfo
  {volume} {108}},\ \bibinfo {pages} {133402} (\bibinfo {year}
  {2012})}\BibitemShut {NoStop}%
\bibitem [{\citenamefont {Platzman}\ and\ \citenamefont
  {Mills}(1994)}]{Platzman94}%
  \BibitemOpen
  \bibfield  {author} {\bibinfo {author} {\bibfnamefont {P.~M.}\ \bibnamefont
  {Platzman}}\ and\ \bibinfo {author} {\bibfnamefont {A.~P.}\ \bibnamefont
  {Mills}},\ }\href {\doibase 10.1103/PhysRevB.49.454} {\bibfield  {journal}
  {\bibinfo  {journal} {Phys. Rev. B}\ }\textbf {\bibinfo {volume} {49}},\
  \bibinfo {pages} {454} (\bibinfo {year} {1994})}\BibitemShut {NoStop}%
\bibitem [{\citenamefont {{Mills, Jr.}}(2007)}]{Mills07}%
  \BibitemOpen
  \bibfield  {author} {\bibinfo {author} {\bibfnamefont {A.~P.}\ \bibnamefont
  {{Mills, Jr.}}},\ }\href {\doibase 10.1016/j.radphyschem.2006.03.009}
  {\bibfield  {journal} {\bibinfo  {journal} {Rad. Phys. Chem.}\ }\textbf
  {\bibinfo {volume} {76}},\ \bibinfo {pages} {76} (\bibinfo {year}
  {2007})}\BibitemShut {NoStop}%
\bibitem [{\citenamefont {Cassidy}\ and\ \citenamefont {{Mills,
  Jr.}}(2007{\natexlab{b}})}]{Cassidy07a}%
  \BibitemOpen
  \bibfield  {author} {\bibinfo {author} {\bibfnamefont {D.~B.}\ \bibnamefont
  {Cassidy}}\ and\ \bibinfo {author} {\bibfnamefont {A.~P.}\ \bibnamefont
  {{Mills, Jr.}}},\ }\href {\doibase 10.1002/pssc.200675760} {\bibfield
  {journal} {\bibinfo  {journal} {Physica Status Solidi (c)}\ }\textbf
  {\bibinfo {volume} {4}},\ \bibinfo {pages} {3419} (\bibinfo {year}
  {2007}{\natexlab{b}})}\BibitemShut {NoStop}%
\bibitem [{\citenamefont {Cassidy}\ and\ \citenamefont
  {Hogan}(2014)}]{Cassidy14}%
  \BibitemOpen
  \bibfield  {author} {\bibinfo {author} {\bibfnamefont {D.~B.}\ \bibnamefont
  {Cassidy}}\ and\ \bibinfo {author} {\bibfnamefont {S.~D.}\ \bibnamefont
  {Hogan}},\ }\href {\doibase 10.1142/S2010194514602592} {\bibfield  {journal}
  {\bibinfo  {journal} {Int. J. Mod. Phys.: Conf. Ser.}\ }\textbf {\bibinfo
  {volume} {30}},\ \bibinfo {pages} {1460259} (\bibinfo {year}
  {2014})}\BibitemShut {NoStop}%
\bibitem [{\citenamefont {Swann}\ \emph {et~al.}(2016)\citenamefont {Swann},
  \citenamefont {Cassidy}, \citenamefont {Deller},\ and\ \citenamefont
  {Gribakin}}]{Swann16}%
  \BibitemOpen
  \bibfield  {author} {\bibinfo {author} {\bibfnamefont {A.~R.}\ \bibnamefont
  {Swann}}, \bibinfo {author} {\bibfnamefont {D.~B.}\ \bibnamefont {Cassidy}},
  \bibinfo {author} {\bibfnamefont {A.}~\bibnamefont {Deller}}, \ and\ \bibinfo
  {author} {\bibfnamefont {G.~F.}\ \bibnamefont {Gribakin}},\ }\href {\doibase
  10.1103/PhysRevA.93.052712} {\bibfield  {journal} {\bibinfo  {journal} {Phys.
  Rev. A}\ }\textbf {\bibinfo {volume} {93}},\ \bibinfo {pages} {052712}
  (\bibinfo {year} {2016})}\BibitemShut {NoStop}%
\bibitem [{\citenamefont {Brawley}\ \emph {et~al.}(2010)\citenamefont
  {Brawley}, \citenamefont {Armitage}, \citenamefont {Beale}, \citenamefont
  {Leslie}, \citenamefont {Williams},\ and\ \citenamefont
  {Laricchia}}]{Brawley10}%
  \BibitemOpen
  \bibfield  {author} {\bibinfo {author} {\bibfnamefont {S.~J.}\ \bibnamefont
  {Brawley}}, \bibinfo {author} {\bibfnamefont {S.~P.}\ \bibnamefont
  {Armitage}}, \bibinfo {author} {\bibfnamefont {J.}~\bibnamefont {Beale}},
  \bibinfo {author} {\bibfnamefont {D.~E.}\ \bibnamefont {Leslie}}, \bibinfo
  {author} {\bibfnamefont {A.~I.}\ \bibnamefont {Williams}}, \ and\ \bibinfo
  {author} {\bibfnamefont {G.}~\bibnamefont {Laricchia}},\ }\href {\doibase
  10.1126/science.1192322} {\bibfield  {journal} {\bibinfo  {journal}
  {Science}\ }\textbf {\bibinfo {volume} {330}},\ \bibinfo {pages} {789}
  (\bibinfo {year} {2010})}\BibitemShut {NoStop}%
\bibitem [{\citenamefont {Brawley}\ \emph {et~al.}(2012)\citenamefont
  {Brawley}, \citenamefont {Williams}, \citenamefont {Shipman},\ and\
  \citenamefont {Laricchia}}]{Brawley12}%
  \BibitemOpen
  \bibfield  {author} {\bibinfo {author} {\bibfnamefont {S.~J.}\ \bibnamefont
  {Brawley}}, \bibinfo {author} {\bibfnamefont {A.~I.}\ \bibnamefont
  {Williams}}, \bibinfo {author} {\bibfnamefont {M.}~\bibnamefont {Shipman}}, \
  and\ \bibinfo {author} {\bibfnamefont {G.}~\bibnamefont {Laricchia}},\ }\href
  {\doibase 10.1088/1742-6596/388/1/012018} {\bibfield  {journal} {\bibinfo
  {journal} {J. Phys.: Conf. Ser.}\ }\textbf {\bibinfo {volume} {388}},\
  \bibinfo {pages} {012018} (\bibinfo {year} {2012})}\BibitemShut {NoStop}%
\bibitem [{\citenamefont {Fabrikant}\ and\ \citenamefont
  {Gribakin}(2014{\natexlab{a}})}]{Fabrikant14a}%
  \BibitemOpen
  \bibfield  {author} {\bibinfo {author} {\bibfnamefont {I.~I.}\ \bibnamefont
  {Fabrikant}}\ and\ \bibinfo {author} {\bibfnamefont {G.~F.}\ \bibnamefont
  {Gribakin}},\ }\href {\doibase 10.1103/PhysRevLett.112.243201} {\bibfield
  {journal} {\bibinfo  {journal} {Phys. Rev. Lett.}\ }\textbf {\bibinfo
  {volume} {112}},\ \bibinfo {pages} {243201} (\bibinfo {year}
  {2014}{\natexlab{a}})}\BibitemShut {NoStop}%
\bibitem [{\citenamefont {Brawley}\ \emph {et~al.}(2015)\citenamefont
  {Brawley}, \citenamefont {Fayer}, \citenamefont {Shipman},\ and\
  \citenamefont {Laricchia}}]{Brawley15}%
  \BibitemOpen
  \bibfield  {author} {\bibinfo {author} {\bibfnamefont {S.~J.}\ \bibnamefont
  {Brawley}}, \bibinfo {author} {\bibfnamefont {S.~E.}\ \bibnamefont {Fayer}},
  \bibinfo {author} {\bibfnamefont {M.}~\bibnamefont {Shipman}}, \ and\
  \bibinfo {author} {\bibfnamefont {G.}~\bibnamefont {Laricchia}},\ }\href
  {\doibase 10.1103/PhysRevLett.115.223201} {\bibfield  {journal} {\bibinfo
  {journal} {Phys. Rev. Lett.}\ }\textbf {\bibinfo {volume} {115}},\ \bibinfo
  {pages} {223201} (\bibinfo {year} {2015})}\BibitemShut {NoStop}%
\bibitem [{\citenamefont {Fabrikant}\ and\ \citenamefont
  {Gribakin}(2014{\natexlab{b}})}]{Fabrikant14}%
  \BibitemOpen
  \bibfield  {author} {\bibinfo {author} {\bibfnamefont {I.~I.}\ \bibnamefont
  {Fabrikant}}\ and\ \bibinfo {author} {\bibfnamefont {G.~F.}\ \bibnamefont
  {Gribakin}},\ }\href {\doibase 10.1103/PhysRevA.90.052717} {\bibfield
  {journal} {\bibinfo  {journal} {Phys. Rev. A}\ }\textbf {\bibinfo {volume}
  {90}},\ \bibinfo {pages} {052717} (\bibinfo {year} {2014}{\natexlab{b}})},\
  \bibinfo {note} {note that Figs. 3 and 8 have been accidentally swapped; Fig.
  3 shows the Ps phase shifts for Ar, and Fig. 8 for Kr.}\BibitemShut {Stop}%
\bibitem [{\citenamefont {Gribakin}\ \emph {et~al.}(2016)\citenamefont
  {Gribakin}, \citenamefont {Swann}, \citenamefont {Wilde},\ and\ \citenamefont
  {Fabrikant}}]{Gribakin16}%
  \BibitemOpen
  \bibfield  {author} {\bibinfo {author} {\bibfnamefont {G.~F.}\ \bibnamefont
  {Gribakin}}, \bibinfo {author} {\bibfnamefont {A.~R.}\ \bibnamefont {Swann}},
  \bibinfo {author} {\bibfnamefont {R.~S.}\ \bibnamefont {Wilde}}, \ and\
  \bibinfo {author} {\bibfnamefont {I.~I.}\ \bibnamefont {Fabrikant}},\ }\href
  {\doibase 10.1088/0953-4075/49/6/064004} {\bibfield  {journal} {\bibinfo
  {journal} {J. Phys. B}\ }\textbf {\bibinfo {volume} {49}},\ \bibinfo {pages}
  {064004} (\bibinfo {year} {2016})}\BibitemShut {NoStop}%
\bibitem [{\citenamefont {Zafar}\ \emph {et~al.}(1996)\citenamefont {Zafar},
  \citenamefont {Laricchia}, \citenamefont {Charlton},\ and\ \citenamefont
  {Garner}}]{Zafar96}%
  \BibitemOpen
  \bibfield  {author} {\bibinfo {author} {\bibfnamefont {N.}~\bibnamefont
  {Zafar}}, \bibinfo {author} {\bibfnamefont {G.}~\bibnamefont {Laricchia}},
  \bibinfo {author} {\bibfnamefont {M.}~\bibnamefont {Charlton}}, \ and\
  \bibinfo {author} {\bibfnamefont {A.}~\bibnamefont {Garner}},\ }\href
  {\doibase 10.1103/PhysRevLett.76.1595} {\bibfield  {journal} {\bibinfo
  {journal} {Phys. Rev. Lett.}\ }\textbf {\bibinfo {volume} {76}},\ \bibinfo
  {pages} {1595} (\bibinfo {year} {1996})}\BibitemShut {NoStop}%
\bibitem [{\citenamefont {Vallery}\ \emph {et~al.}(2003)\citenamefont
  {Vallery}, \citenamefont {Zitzewitz},\ and\ \citenamefont
  {Gidley}}]{Vallery03}%
  \BibitemOpen
  \bibfield  {author} {\bibinfo {author} {\bibfnamefont {R.~S.}\ \bibnamefont
  {Vallery}}, \bibinfo {author} {\bibfnamefont {P.~W.}\ \bibnamefont
  {Zitzewitz}}, \ and\ \bibinfo {author} {\bibfnamefont {D.~W.}\ \bibnamefont
  {Gidley}},\ }\href {\doibase 10.1103/PhysRevLett.90.203402} {\bibfield
  {journal} {\bibinfo  {journal} {Phys. Rev. Lett.}\ }\textbf {\bibinfo
  {volume} {90}},\ \bibinfo {pages} {203402} (\bibinfo {year}
  {2003})}\BibitemShut {NoStop}%
\bibitem [{\citenamefont {Massey}\ and\ \citenamefont {Mohr}(1954)}]{Massey54}%
  \BibitemOpen
  \bibfield  {author} {\bibinfo {author} {\bibfnamefont {H.~S.~W.}\
  \bibnamefont {Massey}}\ and\ \bibinfo {author} {\bibfnamefont {C.~B.~O.}\
  \bibnamefont {Mohr}},\ }\href {\doibase 10.1088/0370-1298/67/8/306}
  {\bibfield  {journal} {\bibinfo  {journal} {Proc. Phys. Soc. A}\ }\textbf
  {\bibinfo {volume} {67}},\ \bibinfo {pages} {695} (\bibinfo {year}
  {1954})}\BibitemShut {NoStop}%
\bibitem [{\citenamefont {Fraser}(1962)}]{Fraser62}%
  \BibitemOpen
  \bibfield  {author} {\bibinfo {author} {\bibfnamefont {P.~A.}\ \bibnamefont
  {Fraser}},\ }\href {\doibase 10.1088/0370-1328/79/4/307} {\bibfield
  {journal} {\bibinfo  {journal} {Proc. Phys. Soc. London}\ }\textbf {\bibinfo
  {volume} {79}},\ \bibinfo {pages} {721} (\bibinfo {year} {1962})}\BibitemShut
  {NoStop}%
\bibitem [{\citenamefont {Fraser}(1968)}]{Fraser68}%
  \BibitemOpen
  \bibfield  {author} {\bibinfo {author} {\bibfnamefont {P.~A.}\ \bibnamefont
  {Fraser}},\ }\href {\doibase 10.1088/0022-3700/1/5/135} {\bibfield  {journal}
  {\bibinfo  {journal} {J. Phys. B}\ }\textbf {\bibinfo {volume} {1}},\
  \bibinfo {pages} {1006} (\bibinfo {year} {1968})}\BibitemShut {NoStop}%
\bibitem [{\citenamefont {Fraser}\ and\ \citenamefont
  {Kraidy}(1966)}]{Fraser66}%
  \BibitemOpen
  \bibfield  {author} {\bibinfo {author} {\bibfnamefont {P.~A.}\ \bibnamefont
  {Fraser}}\ and\ \bibinfo {author} {\bibfnamefont {M.}~\bibnamefont
  {Kraidy}},\ }\href {\doibase 10.1088/0370-1328/89/3/309} {\bibfield
  {journal} {\bibinfo  {journal} {Proc. Phys. Soc. London}\ }\textbf {\bibinfo
  {volume} {89}},\ \bibinfo {pages} {533} (\bibinfo {year} {1966})}\BibitemShut
  {NoStop}%
\bibitem [{\citenamefont {Barker}\ and\ \citenamefont
  {Bransden}(1968)}]{Barker68}%
  \BibitemOpen
  \bibfield  {author} {\bibinfo {author} {\bibfnamefont {M.~I.}\ \bibnamefont
  {Barker}}\ and\ \bibinfo {author} {\bibfnamefont {B.~H.}\ \bibnamefont
  {Bransden}},\ }\href {\doibase 10.1088/0022-3700/1/6/314} {\bibfield
  {journal} {\bibinfo  {journal} {J. Phys. B}\ }\textbf {\bibinfo {volume}
  {1}},\ \bibinfo {pages} {1109} (\bibinfo {year} {1968})}\BibitemShut
  {NoStop}%
\bibitem [{\citenamefont {Barker}\ and\ \citenamefont
  {Bransden}(1969)}]{Barker69}%
  \BibitemOpen
  \bibfield  {author} {\bibinfo {author} {\bibfnamefont {M.~I.}\ \bibnamefont
  {Barker}}\ and\ \bibinfo {author} {\bibfnamefont {B.~H.}\ \bibnamefont
  {Bransden}},\ }\href {\doibase 10.1088/0022-3700/2/6/515} {\bibfield
  {journal} {\bibinfo  {journal} {J. Phys. B}\ }\textbf {\bibinfo {volume}
  {2}},\ \bibinfo {pages} {730} (\bibinfo {year} {1969})}\BibitemShut {NoStop}%
\bibitem [{\citenamefont {Drachman}\ and\ \citenamefont
  {Houston}(1970)}]{Drachman70}%
  \BibitemOpen
  \bibfield  {author} {\bibinfo {author} {\bibfnamefont {R.~J.}\ \bibnamefont
  {Drachman}}\ and\ \bibinfo {author} {\bibfnamefont {S.~K.}\ \bibnamefont
  {Houston}},\ }\href {\doibase 10.1088/0022-3700/3/12/010} {\bibfield
  {journal} {\bibinfo  {journal} {J. Phys. B}\ }\textbf {\bibinfo {volume}
  {3}},\ \bibinfo {pages} {1657} (\bibinfo {year} {1970})}\BibitemShut
  {NoStop}%
\bibitem [{\citenamefont {McAlinden}\ \emph {et~al.}(1996)\citenamefont
  {McAlinden}, \citenamefont {McDonald},\ and\ \citenamefont
  {Walters}}]{McAlinden96}%
  \BibitemOpen
  \bibfield  {author} {\bibinfo {author} {\bibfnamefont {M.~T.}\ \bibnamefont
  {McAlinden}}, \bibinfo {author} {\bibfnamefont {F.~G. R.~S.}\ \bibnamefont
  {McDonald}}, \ and\ \bibinfo {author} {\bibfnamefont {H.~R.~J.}\ \bibnamefont
  {Walters}},\ }\href {\doibase 10.1139/p96-062} {\bibfield  {journal}
  {\bibinfo  {journal} {Can. J. Phys.}\ }\textbf {\bibinfo {volume} {74}},\
  \bibinfo {pages} {434} (\bibinfo {year} {1996})}\BibitemShut {NoStop}%
\bibitem [{\citenamefont {Sarkar}\ and\ \citenamefont
  {Ghosh}(1997)}]{Sarkar97}%
  \BibitemOpen
  \bibfield  {author} {\bibinfo {author} {\bibfnamefont {N.~K.}\ \bibnamefont
  {Sarkar}}\ and\ \bibinfo {author} {\bibfnamefont {A.~S.}\ \bibnamefont
  {Ghosh}},\ }\href {\doibase 10.1088/0953-4075/30/20/019} {\bibfield
  {journal} {\bibinfo  {journal} {J. Phys. B}\ }\textbf {\bibinfo {volume}
  {30}},\ \bibinfo {pages} {4591} (\bibinfo {year} {1997})}\BibitemShut
  {NoStop}%
\bibitem [{\citenamefont {Biswas}\ and\ \citenamefont
  {Adhikari}(1999)}]{Biswas99}%
  \BibitemOpen
  \bibfield  {author} {\bibinfo {author} {\bibfnamefont {P.~K.}\ \bibnamefont
  {Biswas}}\ and\ \bibinfo {author} {\bibfnamefont {S.~K.}\ \bibnamefont
  {Adhikari}},\ }\href {\doibase 10.1103/PhysRevA.59.363} {\bibfield  {journal}
  {\bibinfo  {journal} {Phys. Rev. A}\ }\textbf {\bibinfo {volume} {59}},\
  \bibinfo {pages} {363} (\bibinfo {year} {1999})}\BibitemShut {NoStop}%
\bibitem [{\citenamefont {Sarkar}\ \emph {et~al.}(1999)\citenamefont {Sarkar},
  \citenamefont {Chaudhury},\ and\ \citenamefont {Ghosh}}]{Sarkar99}%
  \BibitemOpen
  \bibfield  {author} {\bibinfo {author} {\bibfnamefont {N.~K.}\ \bibnamefont
  {Sarkar}}, \bibinfo {author} {\bibfnamefont {P.}~\bibnamefont {Chaudhury}}, \
  and\ \bibinfo {author} {\bibfnamefont {A.~S.}\ \bibnamefont {Ghosh}},\ }\href
  {\doibase 10.1088/0953-4075/32/7/009} {\bibfield  {journal} {\bibinfo
  {journal} {J. Phys. B}\ }\textbf {\bibinfo {volume} {32}},\ \bibinfo {pages}
  {1657} (\bibinfo {year} {1999})}\BibitemShut {NoStop}%
\bibitem [{\citenamefont {Blackwood}\ \emph {et~al.}(1999)\citenamefont
  {Blackwood}, \citenamefont {Campbell}, \citenamefont {McAlinden},\ and\
  \citenamefont {Walters}}]{Blackwood99}%
  \BibitemOpen
  \bibfield  {author} {\bibinfo {author} {\bibfnamefont {J.~E.}\ \bibnamefont
  {Blackwood}}, \bibinfo {author} {\bibfnamefont {C.~P.}\ \bibnamefont
  {Campbell}}, \bibinfo {author} {\bibfnamefont {M.~T.}\ \bibnamefont
  {McAlinden}}, \ and\ \bibinfo {author} {\bibfnamefont {H.~R.~J.}\
  \bibnamefont {Walters}},\ }\href {\doibase 10.1103/PhysRevA.60.4454}
  {\bibfield  {journal} {\bibinfo  {journal} {Phys. Rev. A}\ }\textbf {\bibinfo
  {volume} {60}},\ \bibinfo {pages} {4454} (\bibinfo {year}
  {1999})}\BibitemShut {NoStop}%
\bibitem [{\citenamefont {Adhikari}(2000)}]{Adhikari00}%
  \BibitemOpen
  \bibfield  {author} {\bibinfo {author} {\bibfnamefont {S.~K.}\ \bibnamefont
  {Adhikari}},\ }\href {\doibase 10.1103/PhysRevA.62.062708} {\bibfield
  {journal} {\bibinfo  {journal} {Phys. Rev. A}\ }\textbf {\bibinfo {volume}
  {62}},\ \bibinfo {pages} {062708} (\bibinfo {year} {2000})}\BibitemShut
  {NoStop}%
\bibitem [{\citenamefont {Ghosh}\ \emph {et~al.}(2001)\citenamefont {Ghosh},
  \citenamefont {Basu}, \citenamefont {Mukherjee},\ and\ \citenamefont
  {Sinha}}]{Ghosh01}%
  \BibitemOpen
  \bibfield  {author} {\bibinfo {author} {\bibfnamefont {A.~S.}\ \bibnamefont
  {Ghosh}}, \bibinfo {author} {\bibfnamefont {A.}~\bibnamefont {Basu}},
  \bibinfo {author} {\bibfnamefont {T.}~\bibnamefont {Mukherjee}}, \ and\
  \bibinfo {author} {\bibfnamefont {P.~K.}\ \bibnamefont {Sinha}},\ }\href
  {\doibase 10.1103/PhysRevA.63.042706} {\bibfield  {journal} {\bibinfo
  {journal} {Phys. Rev. A}\ }\textbf {\bibinfo {volume} {63}},\ \bibinfo
  {pages} {042706} (\bibinfo {year} {2001})}\BibitemShut {NoStop}%
\bibitem [{\citenamefont {Basu}\ \emph {et~al.}(2001)\citenamefont {Basu},
  \citenamefont {Sinha},\ and\ \citenamefont {Ghosh}}]{Basu01}%
  \BibitemOpen
  \bibfield  {author} {\bibinfo {author} {\bibfnamefont {A.}~\bibnamefont
  {Basu}}, \bibinfo {author} {\bibfnamefont {P.~K.}\ \bibnamefont {Sinha}}, \
  and\ \bibinfo {author} {\bibfnamefont {A.~S.}\ \bibnamefont {Ghosh}},\ }\href
  {\doibase 10.1103/PhysRevA.63.052503} {\bibfield  {journal} {\bibinfo
  {journal} {Phys. Rev. A}\ }\textbf {\bibinfo {volume} {63}},\ \bibinfo
  {pages} {052503} (\bibinfo {year} {2001})}\BibitemShut {NoStop}%
\bibitem [{\citenamefont {Adhikari}(2001)}]{Adhikari01}%
  \BibitemOpen
  \bibfield  {author} {\bibinfo {author} {\bibfnamefont {S.~K.}\ \bibnamefont
  {Adhikari}},\ }\href {\doibase 10.1103/PhysRevA.64.022702} {\bibfield
  {journal} {\bibinfo  {journal} {Phys. Rev. A}\ }\textbf {\bibinfo {volume}
  {64}},\ \bibinfo {pages} {022702} (\bibinfo {year} {2001})}\BibitemShut
  {NoStop}%
\bibitem [{\citenamefont {Ivanov}\ \emph {et~al.}(2001)\citenamefont {Ivanov},
  \citenamefont {Mitroy},\ and\ \citenamefont {Varga}}]{Ivanov01}%
  \BibitemOpen
  \bibfield  {author} {\bibinfo {author} {\bibfnamefont {I.~A.}\ \bibnamefont
  {Ivanov}}, \bibinfo {author} {\bibfnamefont {J.}~\bibnamefont {Mitroy}}, \
  and\ \bibinfo {author} {\bibfnamefont {K.}~\bibnamefont {Varga}},\ }\href
  {\doibase 10.1103/PhysRevLett.87.063201} {\bibfield  {journal} {\bibinfo
  {journal} {Phys. Rev. Lett.}\ }\textbf {\bibinfo {volume} {87}},\ \bibinfo
  {pages} {063201} (\bibinfo {year} {2001})}\BibitemShut {NoStop}%
\bibitem [{\citenamefont {Mitroy}\ and\ \citenamefont
  {Ivanov}(2001)}]{Mitroy01}%
  \BibitemOpen
  \bibfield  {author} {\bibinfo {author} {\bibfnamefont {J.}~\bibnamefont
  {Mitroy}}\ and\ \bibinfo {author} {\bibfnamefont {I.~A.}\ \bibnamefont
  {Ivanov}},\ }\href {\doibase 10.1103/PhysRevA.65.012509} {\bibfield
  {journal} {\bibinfo  {journal} {Phys. Rev. A}\ }\textbf {\bibinfo {volume}
  {65}},\ \bibinfo {pages} {012509} (\bibinfo {year} {2001})}\BibitemShut
  {NoStop}%
\bibitem [{\citenamefont {Chiesa}\ \emph {et~al.}(2002)\citenamefont {Chiesa},
  \citenamefont {Mella},\ and\ \citenamefont {Morosi}}]{Chiesa02}%
  \BibitemOpen
  \bibfield  {author} {\bibinfo {author} {\bibfnamefont {S.}~\bibnamefont
  {Chiesa}}, \bibinfo {author} {\bibfnamefont {M.}~\bibnamefont {Mella}}, \
  and\ \bibinfo {author} {\bibfnamefont {G.}~\bibnamefont {Morosi}},\ }\href
  {\doibase 10.1103/PhysRevA.66.042502} {\bibfield  {journal} {\bibinfo
  {journal} {Phys. Rev. A}\ }\textbf {\bibinfo {volume} {66}},\ \bibinfo
  {pages} {042502} (\bibinfo {year} {2002})}\BibitemShut {NoStop}%
\bibitem [{\citenamefont {DiRenzi}\ and\ \citenamefont
  {Drachman}(2003)}]{DiRenzi03}%
  \BibitemOpen
  \bibfield  {author} {\bibinfo {author} {\bibfnamefont {J.}~\bibnamefont
  {DiRenzi}}\ and\ \bibinfo {author} {\bibfnamefont {R.~J.}\ \bibnamefont
  {Drachman}},\ }\href {\doibase 10.1088/0953-4075/36/12/302} {\bibfield
  {journal} {\bibinfo  {journal} {J. Phys. B}\ }\textbf {\bibinfo {volume}
  {36}},\ \bibinfo {pages} {2409} (\bibinfo {year} {2003})}\BibitemShut
  {NoStop}%
\bibitem [{\citenamefont {Walters}\ \emph {et~al.}(2004)\citenamefont
  {Walters}, \citenamefont {Yu}, \citenamefont {Sahoo},\ and\ \citenamefont
  {Gilmore}}]{Walters04}%
  \BibitemOpen
  \bibfield  {author} {\bibinfo {author} {\bibfnamefont {H.~R.~J.}\
  \bibnamefont {Walters}}, \bibinfo {author} {\bibfnamefont {A.~C.~H.}\
  \bibnamefont {Yu}}, \bibinfo {author} {\bibfnamefont {S.}~\bibnamefont
  {Sahoo}}, \ and\ \bibinfo {author} {\bibfnamefont {S.}~\bibnamefont
  {Gilmore}},\ }\href {\doibase 10.1016/j.nimb.2004.03.047} {\bibfield
  {journal} {\bibinfo  {journal} {Nucl. Instrum. Methods B}\ }\textbf {\bibinfo
  {volume} {221}},\ \bibinfo {pages} {149} (\bibinfo {year}
  {2004})}\BibitemShut {NoStop}%
\bibitem [{\citenamefont {DiRenzi}\ and\ \citenamefont
  {Drachman}(2013)}]{DiRenzi13}%
  \BibitemOpen
  \bibfield  {author} {\bibinfo {author} {\bibfnamefont {J.}~\bibnamefont
  {DiRenzi}}\ and\ \bibinfo {author} {\bibfnamefont {R.~J.}\ \bibnamefont
  {Drachman}},\ }\href {\doibase 10.1139/cjp-2012-0430} {\bibfield  {journal}
  {\bibinfo  {journal} {Can. J. Phys.}\ }\textbf {\bibinfo {volume} {91}},\
  \bibinfo {pages} {188} (\bibinfo {year} {2013})}\BibitemShut {NoStop}%
\bibitem [{\citenamefont {Biswas}\ and\ \citenamefont
  {Adhikari}(2000)}]{Biswas00}%
  \BibitemOpen
  \bibfield  {author} {\bibinfo {author} {\bibfnamefont {P.~K.}\ \bibnamefont
  {Biswas}}\ and\ \bibinfo {author} {\bibfnamefont {S.~K.}\ \bibnamefont
  {Adhikari}},\ }\href {\doibase 10.1016/S0009-2614(99)01354-8} {\bibfield
  {journal} {\bibinfo  {journal} {Chem. Phys. Lett.}\ }\textbf {\bibinfo
  {volume} {317}},\ \bibinfo {pages} {129} (\bibinfo {year}
  {2000})}\BibitemShut {NoStop}%
\bibitem [{\citenamefont {Mitroy}\ and\ \citenamefont
  {Bromley}(2003)}]{Mitroy03}%
  \BibitemOpen
  \bibfield  {author} {\bibinfo {author} {\bibfnamefont {J.}~\bibnamefont
  {Mitroy}}\ and\ \bibinfo {author} {\bibfnamefont {M.~W.~J.}\ \bibnamefont
  {Bromley}},\ }\href {\doibase 10.1103/PhysRevA.67.034502} {\bibfield
  {journal} {\bibinfo  {journal} {Phys. Rev. A}\ }\textbf {\bibinfo {volume}
  {67}},\ \bibinfo {pages} {034502} (\bibinfo {year} {2003})}\BibitemShut
  {NoStop}%
\bibitem [{\citenamefont {Blackwood}\ \emph {et~al.}(2002)\citenamefont
  {Blackwood}, \citenamefont {McAlinden},\ and\ \citenamefont
  {Walters}}]{Blackwood02}%
  \BibitemOpen
  \bibfield  {author} {\bibinfo {author} {\bibfnamefont {J.~E.}\ \bibnamefont
  {Blackwood}}, \bibinfo {author} {\bibfnamefont {M.~T.}\ \bibnamefont
  {McAlinden}}, \ and\ \bibinfo {author} {\bibfnamefont {H.~R.~J.}\
  \bibnamefont {Walters}},\ }\href {\doibase 10.1088/0953-4075/35/12/303}
  {\bibfield  {journal} {\bibinfo  {journal} {J. Phys. B}\ }\textbf {\bibinfo
  {volume} {35}},\ \bibinfo {pages} {2661} (\bibinfo {year}
  {2002})}\BibitemShut {NoStop}%
\bibitem [{\citenamefont {Blackwood}\ \emph {et~al.}(2003)\citenamefont
  {Blackwood}, \citenamefont {McAlinden},\ and\ \citenamefont
  {Walters}}]{Blackwood03}%
  \BibitemOpen
  \bibfield  {author} {\bibinfo {author} {\bibfnamefont {J.~E.}\ \bibnamefont
  {Blackwood}}, \bibinfo {author} {\bibfnamefont {M.~T.}\ \bibnamefont
  {McAlinden}}, \ and\ \bibinfo {author} {\bibfnamefont {H.~R.~J.}\
  \bibnamefont {Walters}},\ }\href {\doibase 10.1088/0953-4075/36/4/402}
  {\bibfield  {journal} {\bibinfo  {journal} {J. Phys. B}\ }\textbf {\bibinfo
  {volume} {36}},\ \bibinfo {pages} {797} (\bibinfo {year} {2003})}\BibitemShut
  {NoStop}%
\bibitem [{\citenamefont {Canter}\ \emph {et~al.}(1975)\citenamefont {Canter},
  \citenamefont {McNutt},\ and\ \citenamefont {Roellig}}]{Canter75}%
  \BibitemOpen
  \bibfield  {author} {\bibinfo {author} {\bibfnamefont {K.~F.}\ \bibnamefont
  {Canter}}, \bibinfo {author} {\bibfnamefont {J.~D.}\ \bibnamefont {McNutt}},
  \ and\ \bibinfo {author} {\bibfnamefont {L.~O.}\ \bibnamefont {Roellig}},\
  }\href {\doibase 10.1103/PhysRevA.12.375} {\bibfield  {journal} {\bibinfo
  {journal} {Phys. Rev. A}\ }\textbf {\bibinfo {volume} {12}},\ \bibinfo
  {pages} {375} (\bibinfo {year} {1975})}\BibitemShut {NoStop}%
\bibitem [{\citenamefont {Rytsola}\ \emph {et~al.}(1984)\citenamefont
  {Rytsola}, \citenamefont {Vettenranta},\ and\ \citenamefont
  {Hautojarvi}}]{Rytsola84}%
  \BibitemOpen
  \bibfield  {author} {\bibinfo {author} {\bibfnamefont {K.}~\bibnamefont
  {Rytsola}}, \bibinfo {author} {\bibfnamefont {J.}~\bibnamefont
  {Vettenranta}}, \ and\ \bibinfo {author} {\bibfnamefont {P.}~\bibnamefont
  {Hautojarvi}},\ }\href {\doibase 10.1088/0022-3700/17/16/018} {\bibfield
  {journal} {\bibinfo  {journal} {J. Phys. B}\ }\textbf {\bibinfo {volume}
  {17}},\ \bibinfo {pages} {3359} (\bibinfo {year} {1984})}\BibitemShut
  {NoStop}%
\bibitem [{\citenamefont {Coleman}\ \emph {et~al.}(1994)\citenamefont
  {Coleman}, \citenamefont {Rayner}, \citenamefont {Jacobsen}, \citenamefont
  {Charlton},\ and\ \citenamefont {West}}]{Coleman94}%
  \BibitemOpen
  \bibfield  {author} {\bibinfo {author} {\bibfnamefont {P.~G.}\ \bibnamefont
  {Coleman}}, \bibinfo {author} {\bibfnamefont {S.}~\bibnamefont {Rayner}},
  \bibinfo {author} {\bibfnamefont {F.~M.}\ \bibnamefont {Jacobsen}}, \bibinfo
  {author} {\bibfnamefont {M.}~\bibnamefont {Charlton}}, \ and\ \bibinfo
  {author} {\bibfnamefont {R.~N.}\ \bibnamefont {West}},\ }\href {\doibase
  10.1088/0953-4075/27/5/017} {\bibfield  {journal} {\bibinfo  {journal} {J.
  Phys. B}\ }\textbf {\bibinfo {volume} {27}},\ \bibinfo {pages} {981}
  (\bibinfo {year} {1994})}\BibitemShut {NoStop}%
\bibitem [{\citenamefont {Nagashima}\ \emph
  {et~al.}(1995{\natexlab{a}})\citenamefont {Nagashima}, \citenamefont {Hyodo},
  \citenamefont {Fujiwara},\ and\ \citenamefont {Ichimura}}]{Nagashima95}%
  \BibitemOpen
  \bibfield  {author} {\bibinfo {author} {\bibfnamefont {Y.}~\bibnamefont
  {Nagashima}}, \bibinfo {author} {\bibfnamefont {T.}~\bibnamefont {Hyodo}},
  \bibinfo {author} {\bibfnamefont {K.}~\bibnamefont {Fujiwara}}, \ and\
  \bibinfo {author} {\bibfnamefont {A.}~\bibnamefont {Ichimura}},\ }in\
  \href@noop {} {\emph {\bibinfo {booktitle} {The International Conference on
  the Physics of Electronic and Atomic Collisions, Vancouver, 1995, Abstracts
  of Contributed Papers}}}\ (\bibinfo {year} {1995})\BibitemShut {NoStop}%
\bibitem [{\citenamefont {Garner}\ \emph {et~al.}(1996)\citenamefont {Garner},
  \citenamefont {Laricchia},\ and\ \citenamefont {\"Ozen}}]{Garner96}%
  \BibitemOpen
  \bibfield  {author} {\bibinfo {author} {\bibfnamefont {A.~J.}\ \bibnamefont
  {Garner}}, \bibinfo {author} {\bibfnamefont {G.}~\bibnamefont {Laricchia}}, \
  and\ \bibinfo {author} {\bibfnamefont {A.}~\bibnamefont {\"Ozen}},\ }\href
  {\doibase 10.1088/0953-4075/29/23/035} {\bibfield  {journal} {\bibinfo
  {journal} {J. Phys. B}\ }\textbf {\bibinfo {volume} {29}},\ \bibinfo {pages}
  {5961} (\bibinfo {year} {1996})}\BibitemShut {NoStop}%
\bibitem [{\citenamefont {Skalsey}\ \emph {et~al.}(1998)\citenamefont
  {Skalsey}, \citenamefont {Engbrecht}, \citenamefont {Bithell}, \citenamefont
  {Vallery},\ and\ \citenamefont {Gidley}}]{Skalsey98}%
  \BibitemOpen
  \bibfield  {author} {\bibinfo {author} {\bibfnamefont {M.}~\bibnamefont
  {Skalsey}}, \bibinfo {author} {\bibfnamefont {J.~J.}\ \bibnamefont
  {Engbrecht}}, \bibinfo {author} {\bibfnamefont {R.~K.}\ \bibnamefont
  {Bithell}}, \bibinfo {author} {\bibfnamefont {R.~S.}\ \bibnamefont
  {Vallery}}, \ and\ \bibinfo {author} {\bibfnamefont {D.~W.}\ \bibnamefont
  {Gidley}},\ }\href {\doibase 10.1103/PhysRevLett.80.3727} {\bibfield
  {journal} {\bibinfo  {journal} {Phys. Rev. Lett.}\ }\textbf {\bibinfo
  {volume} {80}},\ \bibinfo {pages} {3727} (\bibinfo {year}
  {1998})}\BibitemShut {NoStop}%
\bibitem [{\citenamefont {Nagashima}\ \emph {et~al.}(1998)\citenamefont
  {Nagashima}, \citenamefont {Hyodo}, \citenamefont {Fujiwara},\ and\
  \citenamefont {Ichimura}}]{Nagashima98}%
  \BibitemOpen
  \bibfield  {author} {\bibinfo {author} {\bibfnamefont {Y.}~\bibnamefont
  {Nagashima}}, \bibinfo {author} {\bibfnamefont {T.}~\bibnamefont {Hyodo}},
  \bibinfo {author} {\bibfnamefont {K.}~\bibnamefont {Fujiwara}}, \ and\
  \bibinfo {author} {\bibfnamefont {A.}~\bibnamefont {Ichimura}},\ }\href
  {\doibase 10.1088/0953-4075/31/2/014} {\bibfield  {journal} {\bibinfo
  {journal} {J. Phys. B}\ }\textbf {\bibinfo {volume} {31}},\ \bibinfo {pages}
  {329} (\bibinfo {year} {1998})}\BibitemShut {NoStop}%
\bibitem [{\citenamefont {Garner}\ \emph {et~al.}(1998)\citenamefont {Garner},
  \citenamefont {\"Ozen},\ and\ \citenamefont {Laricchia}}]{Garner98}%
  \BibitemOpen
  \bibfield  {author} {\bibinfo {author} {\bibfnamefont {A.~J.}\ \bibnamefont
  {Garner}}, \bibinfo {author} {\bibfnamefont {A.}~\bibnamefont {\"Ozen}}, \
  and\ \bibinfo {author} {\bibfnamefont {G.}~\bibnamefont {Laricchia}},\ }\href
  {\doibase 10.1016/S0168-583X(98)00218-3} {\bibfield  {journal} {\bibinfo
  {journal} {Nucl. Instrum. Methods B}\ }\textbf {\bibinfo {volume} {143}},\
  \bibinfo {pages} {155} (\bibinfo {year} {1998})}\BibitemShut {NoStop}%
\bibitem [{\citenamefont {Nagashima}\ \emph {et~al.}(2001)\citenamefont
  {Nagashima}, \citenamefont {Saito}, \citenamefont {Shinohara},\ and\
  \citenamefont {Hyodo}}]{Nagashima01}%
  \BibitemOpen
  \bibfield  {author} {\bibinfo {author} {\bibfnamefont {Y.}~\bibnamefont
  {Nagashima}}, \bibinfo {author} {\bibfnamefont {F.}~\bibnamefont {Saito}},
  \bibinfo {author} {\bibfnamefont {N.}~\bibnamefont {Shinohara}}, \ and\
  \bibinfo {author} {\bibfnamefont {T.}~\bibnamefont {Hyodo}},\ }in\ \href@noop
  {} {\emph {\bibinfo {booktitle} {New Directions in Antimatter Chemistry and
  Physics}}},\ \bibinfo {editor} {edited by\ \bibinfo {editor} {\bibfnamefont
  {C.~M.}\ \bibnamefont {Surko}}\ and\ \bibinfo {editor} {\bibfnamefont
  {F.~A.}\ \bibnamefont {Gianturco}}}\ (\bibinfo  {publisher} {Kluwer
  Academic},\ \bibinfo {address} {Dordrecht},\ \bibinfo {year}
  {2001})\BibitemShut {NoStop}%
\bibitem [{\citenamefont {Skalsey}\ \emph {et~al.}(2003)\citenamefont
  {Skalsey}, \citenamefont {Engbrecht}, \citenamefont {Nakamura}, \citenamefont
  {Vallery},\ and\ \citenamefont {Gidley}}]{Skalsey03}%
  \BibitemOpen
  \bibfield  {author} {\bibinfo {author} {\bibfnamefont {M.}~\bibnamefont
  {Skalsey}}, \bibinfo {author} {\bibfnamefont {J.~J.}\ \bibnamefont
  {Engbrecht}}, \bibinfo {author} {\bibfnamefont {C.~M.}\ \bibnamefont
  {Nakamura}}, \bibinfo {author} {\bibfnamefont {R.~S.}\ \bibnamefont
  {Vallery}}, \ and\ \bibinfo {author} {\bibfnamefont {D.~W.}\ \bibnamefont
  {Gidley}},\ }\href {\doibase 10.1103/PhysRevA.67.022504} {\bibfield
  {journal} {\bibinfo  {journal} {Phys. Rev. A}\ }\textbf {\bibinfo {volume}
  {67}},\ \bibinfo {pages} {022504} (\bibinfo {year} {2003})}\BibitemShut
  {NoStop}%
\bibitem [{\citenamefont {Engbrecht}\ \emph {et~al.}(2008)\citenamefont
  {Engbrecht}, \citenamefont {Erickson}, \citenamefont {Johnson}, \citenamefont
  {Kolan}, \citenamefont {Legard}, \citenamefont {Lund}, \citenamefont
  {Nyflot},\ and\ \citenamefont {Paulsen}}]{Engbrecht08}%
  \BibitemOpen
  \bibfield  {author} {\bibinfo {author} {\bibfnamefont {J.~J.}\ \bibnamefont
  {Engbrecht}}, \bibinfo {author} {\bibfnamefont {M.~J.}\ \bibnamefont
  {Erickson}}, \bibinfo {author} {\bibfnamefont {C.~P.}\ \bibnamefont
  {Johnson}}, \bibinfo {author} {\bibfnamefont {A.~J.}\ \bibnamefont {Kolan}},
  \bibinfo {author} {\bibfnamefont {A.~E.}\ \bibnamefont {Legard}}, \bibinfo
  {author} {\bibfnamefont {S.~P.}\ \bibnamefont {Lund}}, \bibinfo {author}
  {\bibfnamefont {M.~J.}\ \bibnamefont {Nyflot}}, \ and\ \bibinfo {author}
  {\bibfnamefont {J.~D.}\ \bibnamefont {Paulsen}},\ }\href {\doibase
  10.1103/PhysRevA.77.012711} {\bibfield  {journal} {\bibinfo  {journal} {Phys.
  Rev. A}\ }\textbf {\bibinfo {volume} {77}},\ \bibinfo {pages} {012711}
  (\bibinfo {year} {2008})}\BibitemShut {NoStop}%
\bibitem [{\citenamefont {Sano}\ \emph {et~al.}(2015)\citenamefont {Sano},
  \citenamefont {Kino}, \citenamefont {Oka},\ and\ \citenamefont
  {Sekine}}]{Sano15}%
  \BibitemOpen
  \bibfield  {author} {\bibinfo {author} {\bibfnamefont {Y.}~\bibnamefont
  {Sano}}, \bibinfo {author} {\bibfnamefont {Y.}~\bibnamefont {Kino}}, \bibinfo
  {author} {\bibfnamefont {T.}~\bibnamefont {Oka}}, \ and\ \bibinfo {author}
  {\bibfnamefont {T.}~\bibnamefont {Sekine}},\ }\href {\doibase
  10.1088/1742-6596/618/1/012010} {\bibfield  {journal} {\bibinfo  {journal}
  {J. Phys.: Conf. Ser.}\ }\textbf {\bibinfo {volume} {618}},\ \bibinfo {pages}
  {012010} (\bibinfo {year} {2015})}\BibitemShut {NoStop}%
\bibitem [{\citenamefont {{de Boor}}(2001)}]{deBoor01}%
  \BibitemOpen
  \bibfield  {author} {\bibinfo {author} {\bibfnamefont {C.}~\bibnamefont {{de
  Boor}}},\ }\href@noop {} {\emph {\bibinfo {title} {A Practical Guide to
  Splines}}},\ \bibinfo {edition} {revised}\ ed.,\ \bibinfo {series} {Applied
  Mathematical Sciences}, Vol.~\bibinfo {volume} {27}\ (\bibinfo  {publisher}
  {Springer},\ \bibinfo {address} {New York},\ \bibinfo {year}
  {2001})\BibitemShut {NoStop}%
\bibitem [{\citenamefont {Sapirstein}\ and\ \citenamefont
  {Johnson}(1996)}]{Sapirstein96}%
  \BibitemOpen
  \bibfield  {author} {\bibinfo {author} {\bibfnamefont {J.}~\bibnamefont
  {Sapirstein}}\ and\ \bibinfo {author} {\bibfnamefont {W.~R.}\ \bibnamefont
  {Johnson}},\ }\href {\doibase 10.1088/0953-4075/29/22/005} {\bibfield
  {journal} {\bibinfo  {journal} {J. Phys. B}\ }\textbf {\bibinfo {volume}
  {29}},\ \bibinfo {pages} {5213} (\bibinfo {year} {1996})}\BibitemShut
  {NoStop}%
\bibitem [{\citenamefont {Bachau}\ \emph {et~al.}(2001)\citenamefont {Bachau},
  \citenamefont {Cormier}, \citenamefont {Decleva}, \citenamefont {Hansen},\
  and\ \citenamefont {Mart\'in}}]{Bachau01}%
  \BibitemOpen
  \bibfield  {author} {\bibinfo {author} {\bibfnamefont {H.}~\bibnamefont
  {Bachau}}, \bibinfo {author} {\bibfnamefont {E.}~\bibnamefont {Cormier}},
  \bibinfo {author} {\bibfnamefont {P.}~\bibnamefont {Decleva}}, \bibinfo
  {author} {\bibfnamefont {J.~E.}\ \bibnamefont {Hansen}}, \ and\ \bibinfo
  {author} {\bibfnamefont {F.}~\bibnamefont {Mart\'in}},\ }\href {\doibase
  10.1088/0034-4885/64/12/205} {\bibfield  {journal} {\bibinfo  {journal} {Rep.
  Prog. Phys.}\ }\textbf {\bibinfo {volume} {64}},\ \bibinfo {pages} {1815}
  (\bibinfo {year} {2001})}\BibitemShut {NoStop}%
\bibitem [{\citenamefont {Brown}\ \emph {et~al.}(2017)\citenamefont {Brown},
  \citenamefont {Prigent}, \citenamefont {Swann},\ and\ \citenamefont
  {Gribakin}}]{Brown17}%
  \BibitemOpen
  \bibfield  {author} {\bibinfo {author} {\bibfnamefont {R.}~\bibnamefont
  {Brown}}, \bibinfo {author} {\bibfnamefont {Q.}~\bibnamefont {Prigent}},
  \bibinfo {author} {\bibfnamefont {A.~R.}\ \bibnamefont {Swann}}, \ and\
  \bibinfo {author} {\bibfnamefont {G.~F.}\ \bibnamefont {Gribakin}},\ }\href
  {\doibase 10.1103/PhysRevA.95.032705} {\bibfield  {journal} {\bibinfo
  {journal} {Phys. Rev. A}\ }\textbf {\bibinfo {volume} {95}},\ \bibinfo
  {pages} {032705} (\bibinfo {year} {2017})}\BibitemShut {NoStop}%
\bibitem [{\citenamefont {Burke}(1977)}]{Burke77}%
  \BibitemOpen
  \bibfield  {author} {\bibinfo {author} {\bibfnamefont {P.~G.}\ \bibnamefont
  {Burke}},\ }\href@noop {} {\emph {\bibinfo {title} {Potential Scattering in
  Atomic Physics}}}\ (\bibinfo  {publisher} {Plenum Press},\ \bibinfo {address}
  {New York},\ \bibinfo {year} {1977})\BibitemShut {NoStop}%
\bibitem [{\citenamefont {Percival}(1957)}]{Percival57}%
  \BibitemOpen
  \bibfield  {author} {\bibinfo {author} {\bibfnamefont {I.~C.}\ \bibnamefont
  {Percival}},\ }\href {\doibase 10.1088/0370-1298/70/7/303} {\bibfield
  {journal} {\bibinfo  {journal} {Proc. Phys. Soc. A}\ }\textbf {\bibinfo
  {volume} {70}},\ \bibinfo {pages} {494} (\bibinfo {year} {1957})}\BibitemShut
  {NoStop}%
\bibitem [{\citenamefont {Alhassid}\ and\ \citenamefont
  {Koonin}(1984)}]{Alhassid84}%
  \BibitemOpen
  \bibfield  {author} {\bibinfo {author} {\bibfnamefont {Y.}~\bibnamefont
  {Alhassid}}\ and\ \bibinfo {author} {\bibfnamefont {S.~E.}\ \bibnamefont
  {Koonin}},\ }\href {\doibase 10.1016/0003-4916(84)90254-9} {\bibfield
  {journal} {\bibinfo  {journal} {Ann. Phys.}\ }\textbf {\bibinfo {volume}
  {155}},\ \bibinfo {pages} {108} (\bibinfo {year} {1984})}\BibitemShut
  {NoStop}%
\bibitem [{\citenamefont {{van Faassen}}\ \emph {et~al.}(2007)\citenamefont
  {{van Faassen}}, \citenamefont {Wasserman}, \citenamefont {Engel},
  \citenamefont {Zhang},\ and\ \citenamefont {Burke}}]{vanFaasen07}%
  \BibitemOpen
  \bibfield  {author} {\bibinfo {author} {\bibfnamefont {M.}~\bibnamefont {{van
  Faassen}}}, \bibinfo {author} {\bibfnamefont {A.}~\bibnamefont {Wasserman}},
  \bibinfo {author} {\bibfnamefont {E.}~\bibnamefont {Engel}}, \bibinfo
  {author} {\bibfnamefont {F.}~\bibnamefont {Zhang}}, \ and\ \bibinfo {author}
  {\bibfnamefont {K.}~\bibnamefont {Burke}},\ }\href {\doibase
  10.1103/PhysRevLett.99.043005} {\bibfield  {journal} {\bibinfo  {journal}
  {Phys. Rev. Lett.}\ }\textbf {\bibinfo {volume} {99}},\ \bibinfo {pages}
  {043005} (\bibinfo {year} {2007})}\BibitemShut {NoStop}%
\bibitem [{\citenamefont {Cheng}\ \emph {et~al.}(2014)\citenamefont {Cheng},
  \citenamefont {Tang}, \citenamefont {Mitroy},\ and\ \citenamefont
  {Safronova}}]{Cheng14}%
  \BibitemOpen
  \bibfield  {author} {\bibinfo {author} {\bibfnamefont {Y.}~\bibnamefont
  {Cheng}}, \bibinfo {author} {\bibfnamefont {L.~Y.}\ \bibnamefont {Tang}},
  \bibinfo {author} {\bibfnamefont {J.}~\bibnamefont {Mitroy}}, \ and\ \bibinfo
  {author} {\bibfnamefont {M.~S.}\ \bibnamefont {Safronova}},\ }\href {\doibase
  10.1103/PhysRevA.89.012701} {\bibfield  {journal} {\bibinfo  {journal} {Phys.
  Rev. A}\ }\textbf {\bibinfo {volume} {89}},\ \bibinfo {pages} {012701}
  (\bibinfo {year} {2014})}\BibitemShut {NoStop}%
\bibitem [{\citenamefont {Ganas}(1972)}]{Ganas72}%
  \BibitemOpen
  \bibfield  {author} {\bibinfo {author} {\bibfnamefont {P.~S.}\ \bibnamefont
  {Ganas}},\ }\href {\doibase 10.1103/PhysRevA.5.1684} {\bibfield  {journal}
  {\bibinfo  {journal} {Phys. Rev. A}\ }\textbf {\bibinfo {volume} {5}},\
  \bibinfo {pages} {1684} (\bibinfo {year} {1972})}\BibitemShut {NoStop}%
\bibitem [{\citenamefont {Swann}\ \emph {et~al.}(2015)\citenamefont {Swann},
  \citenamefont {Ludlow},\ and\ \citenamefont {Gribakin}}]{Swann15}%
  \BibitemOpen
  \bibfield  {author} {\bibinfo {author} {\bibfnamefont {A.~R.}\ \bibnamefont
  {Swann}}, \bibinfo {author} {\bibfnamefont {J.~A.}\ \bibnamefont {Ludlow}}, \
  and\ \bibinfo {author} {\bibfnamefont {G.~F.}\ \bibnamefont {Gribakin}},\
  }\href {\doibase 10.1103/PhysRevA.92.012505} {\bibfield  {journal} {\bibinfo
  {journal} {Phys. Rev. A}\ }\textbf {\bibinfo {volume} {92}},\ \bibinfo
  {pages} {012505} (\bibinfo {year} {2015})}\BibitemShut {NoStop}%
\bibitem [{\citenamefont {Bargmann}(1952)}]{Bargmann52}%
  \BibitemOpen
  \bibfield  {author} {\bibinfo {author} {\bibfnamefont {V.}~\bibnamefont
  {Bargmann}},\ }\href {http://www.ncbi.nlm.nih.gov/pmc/articles/PMC1063691/}
  {\bibfield  {journal} {\bibinfo  {journal} {Proc. Natl. Acad. Sci. USA}\
  }\textbf {\bibinfo {volume} {38}},\ \bibinfo {pages} {961} (\bibinfo {year}
  {1952})}\BibitemShut {NoStop}%
\bibitem [{\citenamefont {Dzuba}\ and\ \citenamefont
  {Gribakin}(1994)}]{Dzuba94}%
  \BibitemOpen
  \bibfield  {author} {\bibinfo {author} {\bibfnamefont {V.~A.}\ \bibnamefont
  {Dzuba}}\ and\ \bibinfo {author} {\bibfnamefont {G.~F.}\ \bibnamefont
  {Gribakin}},\ }\href {\doibase 10.1103/PhysRevA.49.2483} {\bibfield
  {journal} {\bibinfo  {journal} {Phys. Rev. A}\ }\textbf {\bibinfo {volume}
  {49}},\ \bibinfo {pages} {2483} (\bibinfo {year} {1994})}\BibitemShut
  {NoStop}%
\bibitem [{\citenamefont {{Ya. Amusia}}\ and\ \citenamefont
  {Chernysheva}(1997)}]{Amusia97}%
  \BibitemOpen
  \bibfield  {author} {\bibinfo {author} {\bibfnamefont {M.}~\bibnamefont {{Ya.
  Amusia}}}\ and\ \bibinfo {author} {\bibfnamefont {L.~V.}\ \bibnamefont
  {Chernysheva}},\ }\href@noop {} {\emph {\bibinfo {title} {Computation of
  Atomic Processes: A Handbook for the ATOM Programs}}}\ (\bibinfo  {publisher}
  {IOP Publishing},\ \bibinfo {address} {Bristol},\ \bibinfo {year}
  {1997})\BibitemShut {NoStop}%
\bibitem [{\citenamefont {Gribakin}\ and\ \citenamefont
  {Ludlow}(2004)}]{Gribakin04}%
  \BibitemOpen
  \bibfield  {author} {\bibinfo {author} {\bibfnamefont {G.~F.}\ \bibnamefont
  {Gribakin}}\ and\ \bibinfo {author} {\bibfnamefont {J.}~\bibnamefont
  {Ludlow}},\ }\href {\doibase 10.1103/PhysRevA.70.032720} {\bibfield
  {journal} {\bibinfo  {journal} {Phys. Rev. A}\ }\textbf {\bibinfo {volume}
  {70}},\ \bibinfo {pages} {032720} (\bibinfo {year} {2004})}\BibitemShut
  {NoStop}%
\bibitem [{Note1()}]{Note1}%
  \BibitemOpen
  \bibinfo {note} {Compared with the exponential knot sequence, which is
  convenient for atomic calculations, including efficient convergence of
  many-body-theory sums \cite {Gribakin04}, the ``quadratic-linear'' knot
  sequence (\ref {eq:QLknots}) ensures that the Ps states, Eq. (\ref
  {eq:CIexpansion}), are represented accurately both at small distances and
  near the cavity wall.}\BibitemShut {Stop}%
\bibitem [{\citenamefont {Ghosh}\ and\ \citenamefont {Sinha}(2001)}]{Ghosh01a}%
  \BibitemOpen
  \bibfield  {author} {\bibinfo {author} {\bibfnamefont {A.~S.}\ \bibnamefont
  {Ghosh}}\ and\ \bibinfo {author} {\bibfnamefont {P.~K.}\ \bibnamefont
  {Sinha}},\ }in\ \href@noop {} {\emph {\bibinfo {booktitle} {New Directions in
  Antimatter Chemistry and Physics}}},\ \bibinfo {editor} {edited by\ \bibinfo
  {editor} {\bibfnamefont {C.~M.}\ \bibnamefont {Surko}}\ and\ \bibinfo
  {editor} {\bibfnamefont {F.~A.}\ \bibnamefont {Gianturco}}}\ (\bibinfo
  {publisher} {Kluwer Academic},\ \bibinfo {address} {Dordrecht},\ \bibinfo
  {year} {2001})\BibitemShut {NoStop}%
\bibitem [{\citenamefont {Schwerdtfeger}(2006)}]{onlinepolarizabilities}%
  \BibitemOpen
  \bibfield  {author} {\bibinfo {author} {\bibfnamefont {P.}~\bibnamefont
  {Schwerdtfeger}},\ }\enquote {\bibinfo {title} {Atomic static dipole
  polarizabilities},}\ in\ \href@noop {} {\emph {\bibinfo {booktitle}
  {Computational Aspects of Electric Polarizability Calculations: Atoms,
  Molecules and Clusters}}},\ \bibinfo {editor} {edited by\ \bibinfo {editor}
  {\bibfnamefont {G.}~\bibnamefont {Maroulis}}}\ (\bibinfo  {publisher} {IOS
  Press},\ \bibinfo {address} {Amsterdam},\ \bibinfo {year} {2006})\ pp.\
  \bibinfo {pages} {1--32},\ \bibinfo {note} {updated static dipole
  polarizabilities are available as a PDF file from the CTCP website at Massey
  University:
  \url{http://ctcp.massey.ac.nz/dipole-polarizabilities}.}\BibitemShut {Stop}%
\bibitem [{\citenamefont {Sauder}(1968)}]{Sauder68}%
  \BibitemOpen
  \bibfield  {author} {\bibinfo {author} {\bibfnamefont {W.~C.}\ \bibnamefont
  {Sauder}},\ }\href
  {http://nvlpubs.nist.gov/nistpubs/jres/72A/jresv72An1p91_A1b.pdf} {\bibfield
  {journal} {\bibinfo  {journal} {J. Res. Natl. Bur. Stand. Sect. A}\ }\textbf
  {\bibinfo {volume} {72A}},\ \bibinfo {pages} {91} (\bibinfo {year}
  {1968})}\BibitemShut {NoStop}%
\bibitem [{\citenamefont {Nagashima}\ \emph
  {et~al.}(1995{\natexlab{b}})\citenamefont {Nagashima}, \citenamefont
  {Kakimoto}, \citenamefont {Hyodo}, \citenamefont {Fujiwara}, \citenamefont
  {Ichimura}, \citenamefont {Chang}, \citenamefont {Deng}, \citenamefont
  {Akahane}, \citenamefont {Chiba}, \citenamefont {Suzuki}, \citenamefont
  {McKee},\ and\ \citenamefont {Stewart}}]{Nagashima95a}%
  \BibitemOpen
  \bibfield  {author} {\bibinfo {author} {\bibfnamefont {Y.}~\bibnamefont
  {Nagashima}}, \bibinfo {author} {\bibfnamefont {M.}~\bibnamefont {Kakimoto}},
  \bibinfo {author} {\bibfnamefont {T.}~\bibnamefont {Hyodo}}, \bibinfo
  {author} {\bibfnamefont {K.}~\bibnamefont {Fujiwara}}, \bibinfo {author}
  {\bibfnamefont {A.}~\bibnamefont {Ichimura}}, \bibinfo {author}
  {\bibfnamefont {T.}~\bibnamefont {Chang}}, \bibinfo {author} {\bibfnamefont
  {J.}~\bibnamefont {Deng}}, \bibinfo {author} {\bibfnamefont {T.}~\bibnamefont
  {Akahane}}, \bibinfo {author} {\bibfnamefont {T.}~\bibnamefont {Chiba}},
  \bibinfo {author} {\bibfnamefont {K.}~\bibnamefont {Suzuki}}, \bibinfo
  {author} {\bibfnamefont {B.~T.~A.}\ \bibnamefont {McKee}}, \ and\ \bibinfo
  {author} {\bibfnamefont {A.~T.}\ \bibnamefont {Stewart}},\ }\href {\doibase
  10.1103/PhysRevA.52.258} {\bibfield  {journal} {\bibinfo  {journal} {Phys.
  Rev. A}\ }\textbf {\bibinfo {volume} {52}},\ \bibinfo {pages} {258} (\bibinfo
  {year} {1995}{\natexlab{b}})}\BibitemShut {NoStop}%
\bibitem [{\citenamefont {Dzuba}\ \emph {et~al.}(1996)\citenamefont {Dzuba},
  \citenamefont {Flambaum}, \citenamefont {Gribakin},\ and\ \citenamefont
  {King}}]{Dzuba96}%
  \BibitemOpen
  \bibfield  {author} {\bibinfo {author} {\bibfnamefont {V.~A.}\ \bibnamefont
  {Dzuba}}, \bibinfo {author} {\bibfnamefont {V.~V.}\ \bibnamefont {Flambaum}},
  \bibinfo {author} {\bibfnamefont {G.~F.}\ \bibnamefont {Gribakin}}, \ and\
  \bibinfo {author} {\bibfnamefont {W.~A.}\ \bibnamefont {King}},\ }\href
  {\doibase 10.1088/0953-4075/29/14/024} {\bibfield  {journal} {\bibinfo
  {journal} {J. Phys. B}\ }\textbf {\bibinfo {volume} {29}},\ \bibinfo {pages}
  {3151} (\bibinfo {year} {1996})}\BibitemShut {NoStop}%
\bibitem [{\citenamefont {Green}\ \emph {et~al.}(2014)\citenamefont {Green},
  \citenamefont {Ludlow},\ and\ \citenamefont {Gribakin}}]{Green14}%
  \BibitemOpen
  \bibfield  {author} {\bibinfo {author} {\bibfnamefont {D.~G.}\ \bibnamefont
  {Green}}, \bibinfo {author} {\bibfnamefont {J.~A.}\ \bibnamefont {Ludlow}}, \
  and\ \bibinfo {author} {\bibfnamefont {G.~F.}\ \bibnamefont {Gribakin}},\
  }\href {\doibase 10.1103/PhysRevA.90.032712} {\bibfield  {journal} {\bibinfo
  {journal} {Phys. Rev. A}\ }\textbf {\bibinfo {volume} {90}},\ \bibinfo
  {pages} {032712} (\bibinfo {year} {2014})}\BibitemShut {NoStop}%
\bibitem [{\citenamefont {Green}\ and\ \citenamefont
  {Gribakin}(2013)}]{Green13}%
  \BibitemOpen
  \bibfield  {author} {\bibinfo {author} {\bibfnamefont {D.~G.}\ \bibnamefont
  {Green}}\ and\ \bibinfo {author} {\bibfnamefont {G.~F.}\ \bibnamefont
  {Gribakin}},\ }\href {\doibase 10.1103/PhysRevA.88.032708} {\bibfield
  {journal} {\bibinfo  {journal} {Phys. Rev. A}\ }\textbf {\bibinfo {volume}
  {88}},\ \bibinfo {pages} {032708} (\bibinfo {year} {2013})}\BibitemShut
  {NoStop}%
\bibitem [{\citenamefont {Green}\ and\ \citenamefont
  {Gribakin}(2015)}]{Green15}%
  \BibitemOpen
  \bibfield  {author} {\bibinfo {author} {\bibfnamefont {D.~G.}\ \bibnamefont
  {Green}}\ and\ \bibinfo {author} {\bibfnamefont {G.~F.}\ \bibnamefont
  {Gribakin}},\ }\href {\doibase 10.1103/PhysRevLett.114.093201} {\bibfield
  {journal} {\bibinfo  {journal} {Phys. Rev. Lett.}\ }\textbf {\bibinfo
  {volume} {114}},\ \bibinfo {pages} {093201} (\bibinfo {year}
  {2015})}\BibitemShut {NoStop}%
\end{thebibliography}
\end{document}